\documentclass[traditabstract]{aa}
\usepackage{natbib,twoopt,amssymb,lscape,psfig}
\usepackage[breaklinks=true]{hyperref} 
\bibpunct{(}{)}{;}{a}{}{,} 
\newcommandtwoopt{\citeads}[3][][]{\href{http://adsabs.harvard.edu/abs/#3}%
{\citealp[#1][#2]{#3}}} 
\newcommandtwoopt{\citepads}[3][][]{\href{http://adsabs.harvard.edu/abs/#3}%
{\citep[#1][#2]{#3}}} 
\newcommandtwoopt{\citetads}[3][][]{\href{http://adsabs.harvard.edu/abs/#3}%
{\citet[#1][#2]{#3}}} 
\newcommandtwoopt{\citeyearads}[3][][]%
{\href{http://adsabs.harvard.edu/abs/#3}{\citeyear[#1][#2]{#3}}}
\usepackage{graphicx}
\usepackage{longtable}
\usepackage{natbib}
\bibpunct{(}{)}{;}{a}{}{,} 

\usepackage{txfonts}
\newcommand{\suzaku}{{\it Suzaku}}
\newcommand{\xmm}{{\it XMM-Newton}}

\newcommand{\swift}{{\it Swift}}
\newcommand{\integral}{\textit{INTEGRAL}}
\newcommand{\nustar}{\textit{NuSTAR}}

\newcommand{\ms}{$M_{\odot}$}

\newcommand{\lumcgs}{ergs~s$^{-1}$}

\newcommand{\nodata}{...}

\usepackage{xcolor,ulem}

\begin{document}

   \title{Taking a break: paused accretion in the symbiotic binary RT~Cru}

   
\author{A. Pujol \inst{1,2,3} 
\and
G. J. M. Luna \inst{1,2,3,*}
\and
K. Mukai \inst{4,5}
\and
J. L. Sokoloski \inst{6}
\and
N. P. M. Kuin \inst{7}
\and
F. M. Walter \inst{8}
\and
R.~Angeloni \inst{9}
\and
Y. Nikolov \inst{10}
\and
R. Lopes de Oliveira \inst{11,12}
\and
N. E. Nu\~nez \inst{13}
\and
M. Jaque Arancibia \inst{14,15}
\and
T. Palma \inst{16,17}
\and
L. Gramajo \inst{16,17}
}
\institute{CONICET-Instituto de Astronom\'ia y F\'isica del Espacio
Ciudad Universitaria – Pabellon 2
Intendente G\"uiraldes 2160
(C1428EGA) Ciudad Aut\'onoma de Buenos Aires - Argentina
\email{gjmluna@iafe.uba.ar}
\and
Universidad de Buenos Aires, Facultad de Ciencias Exactas y Naturales, Buenos Aires, Argentina. 
\and
Universidad Nacional de Hurlingham, Av. Gdor. Vergara 2222, Villa Tesei, Buenos Aires, Argentina
\and
CRESST and X-ray Astrophysics Laboratory, NASA Goddard Space Flight Center, Greenbelt, MD 20771, USA 
\and
Department of Physics, University of Maryland, Baltimore County, 1000 Hilltop Circle, Baltimore, MD 21250, USA 
\and
Columbia Astrophysics Lab 550 W120th St., 1027 Pupin Hall, MC 5247 Columbia University, New York, New York 10027, USA 
\and
Mullard Space Science Laboratory, University College London, Holmbury St Mary, Dorking, Surrey RH5 6NT, UK 
\and
Department of  Physics \& Astronomy, Stony Brook University, Stony Brook NY 11794-3800, USA
\and
Gemini Observatory / NSF’s NOIRLab, Casilla 603, La Serena, Chile
\and 
Institute of Astronomy and National Astronomical Observatory, Bulgarian Academy of Sciences, 72 Tsarigradsko Chaussee Blvd., 1784, Sofia, Bulgaria 
\and
Departamento de Física, Universidade Federal de Sergipe, Av. Marechal Rondon, S/N, 49100-000, São Cristóvão, SE, Brazil 
\and
Observat\'orio Nacional, Rua Gal. Jos\'e Cristino 77, 20921-400, Rio~de~Janeiro, RJ, Brazil
\and
Instituto de Ciencias Astronómicas, de la Tierra y del Espacio (ICATE-CONICET), Av. España Sur 1512, J5402DSP, San Juan, Argentina
\and
Instituto de Investigaci\'on Multidisciplinar en Ciencia y Tecnolog\'ia, Universidad de La Serena, Av. R. Bitr\'an 1305, La Serena, Chile 
\and
Departamento de Astronom\'ia, Universidad de La Serena, Av. J. Cisternas 1200, La Serena, Chile
\and
Universidad Nacional de C\'ordoba, Observatorio Astron\'omico de C\'ordoba, Laprida 854, 5000 C\'ordoba, Argentina 
\and
Consejo Nacional de Investigaciones Cient\'ificas y T\'ecnicas (CONICET), Godoy Cruz 2290, Ciudad Aut\'onoma de Buenos Aires, Argentina
}

   \date{Received ; accepted }

 
  \abstract
{Symbiotic binaries sometimes hide their symbiotic nature for significant periods of time. There is mounting observational evidence that in those symbiotics that are powered solely by accretion of red-giant's wind material onto a white dwarf, without any quasi-steady shell burning on the surface of the white dwarf, the characteristic emission lines in the optical spectrum can vanish, leaving the semblance of an isolated red giant spectrum. Here we present compelling evidence that this disappearance of optical emission lines from the spectrum of RT Cru during 2019 was due to a decrease in the accretion rate, which we derive by modeling the X-ray spectrum. This drop in accretion rate leads to a lower flux of ionizing photons and thus to faint/absent photoionization emission lines in the optical spectrum. We observed the white dwarf symbiotic RT~Cru with \xmm\ and \swift\ in X-rays and UV and collected ground-based optical spectra and photometry over the last 33 years. This long-term coverage shows that during most of the year 2019, the accretion rate onto the white dwarf was so low, $\dot{M}= (3.2\pm 0.06)\, \times$10$^{-11}$ \ms\,yr$^{-1}$ (d/2.52 kpc)$^2$, that the historically detected hard X-ray emission almost vanished, the UV flux faded by roughly 5 magnitudes, the $U$, $B$ and $V$ flickering amplitude decreased, and the Balmer lines virtually disappeared from January through March 2019. Long-lasting low-accretion episodes as the one reported here may hamper the chances of RT~Cru experiencing nova-type outburst despite the high-mass of the accreting white dwarf.

}


   \maketitle
%

\section{Introduction \label{sec:intro}}

\cite{2013A&A...559A...6L} proposed a definition for a symbiotic system, aiming to be as free as possible from observational biases: ``{\it  a binary in which a red giant transfers enough material to a compact companion to produce an observable signal at any wavelength}''. Among the symbiotic systems, those accreting onto a white dwarf (WD) are dubbed {\it white dwarf symbiotics}. Among the white dwarf symbiotics, those systems with high WD luminosity (L\,$\gtrsim$\,10\,L$_{\odot}$) are powered by nuclear burning on the white dwarf surface, while those with lower WD luminosities are powered by accretion. 
Among the accretion-powered white dwarf symbiotics, one system stands out for its particularly high energy emission: RT~Cru. RT~Cru was first detected in the high energy regime with \integral\ in 2003--2004 \citep{2005ATel..519....1C} at approximately a 3 mCrab level, in 2012 at a 13 mCrab level \citep{2012ATel.3887....1S} and in 2015 at a 6 mCrab level \citep{2015ATel.8448....1S}. \citet{2018A&A...616A..53L} analyzed multi-wavelength data covering the late-2000 to mid-2017 period, when 
two brightening episodes with small amplitude ($\Delta$V $\sim$ 1.5) were observed in optical, with one of them simultaneously covered with \swift/BAT in high energies. The authors suggested that the brightness increases when the accretion rate through the accretion disk rises, but does not rise by the amount necessary to significantly change the optical depth of the disk's innermost region. This behavior stands in contrast to what is observed in most dwarf novae or other symbiotics \citep[e.g. T CrB,][]{2018A&A...619A..61L}, where the accretion rate increased above the amount necessary to change the optical depth of the boundary layer, most likely due to disk instabilities.

In this article, we describe new multiwavelength observations of RT~Cru collected between 1989 and early 2022 that indicate a dramatic fade in the flux at all wavelengths since the beginning of 2019. In the next sections, we describe the observations and the reduction procedure for every dataset analyzed, showing that high dispersion spectra revealed no optical emission lines from January through March 2019,
while the fading was more pronounced and lasted longer in the high energy regime. 
Throughout the article, we assume a distance of 2.52$\pm$0.19 kpc \citep{2021yCat.1352....0B}.
The results of the analysis of X-ray, UV, and optical data are described in Sect. \ref{sec:res}.
In Sect. \ref{sec:disc}, we discuss our conclusions that
the observed fading episode was due to a decrease in the rate of accretion 
through the disk. 

\section{Observations and data analysis \label{sec:obs}}

\begin{figure*}
\includegraphics[scale=0.80]{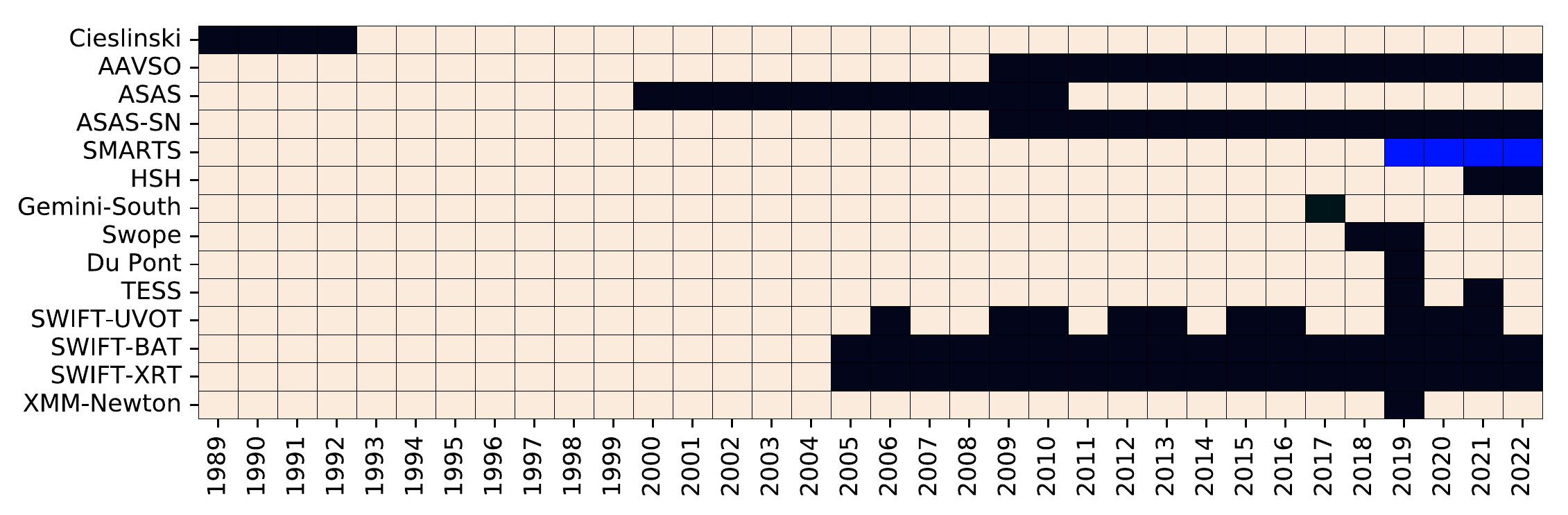}
\caption{Schematic view of the datasets used in this study, where filler black squares highlight the years of the observation obtained from a given instrument, database or literature. Blue squares mark the year of the optical spectroscopic data.
}
\label{fig:obs_log}
\end{figure*}

We observed RT~Cru in X-rays and UV with \xmm\ and \swift\ and obtained optical spectra with the SMARTS 
telescope. Fast optical photometry was obtained with the HSH (Helen Sawyer Hogg), $Gemini$-South, {\it Swope}, {\it Du Pont} and {\it TESS} telescopes. We also collected multi-epoch photometric observations in the V and B bands from the American Association of Variable Star Observers (AAVSO) and the All Sky Automated Survey \citep[ASAS;][]{2002AcA....52..397P}. Figure \ref{fig:obs_log} summarize the datasets described above. 

\begin{figure*}
\includegraphics[scale=1]{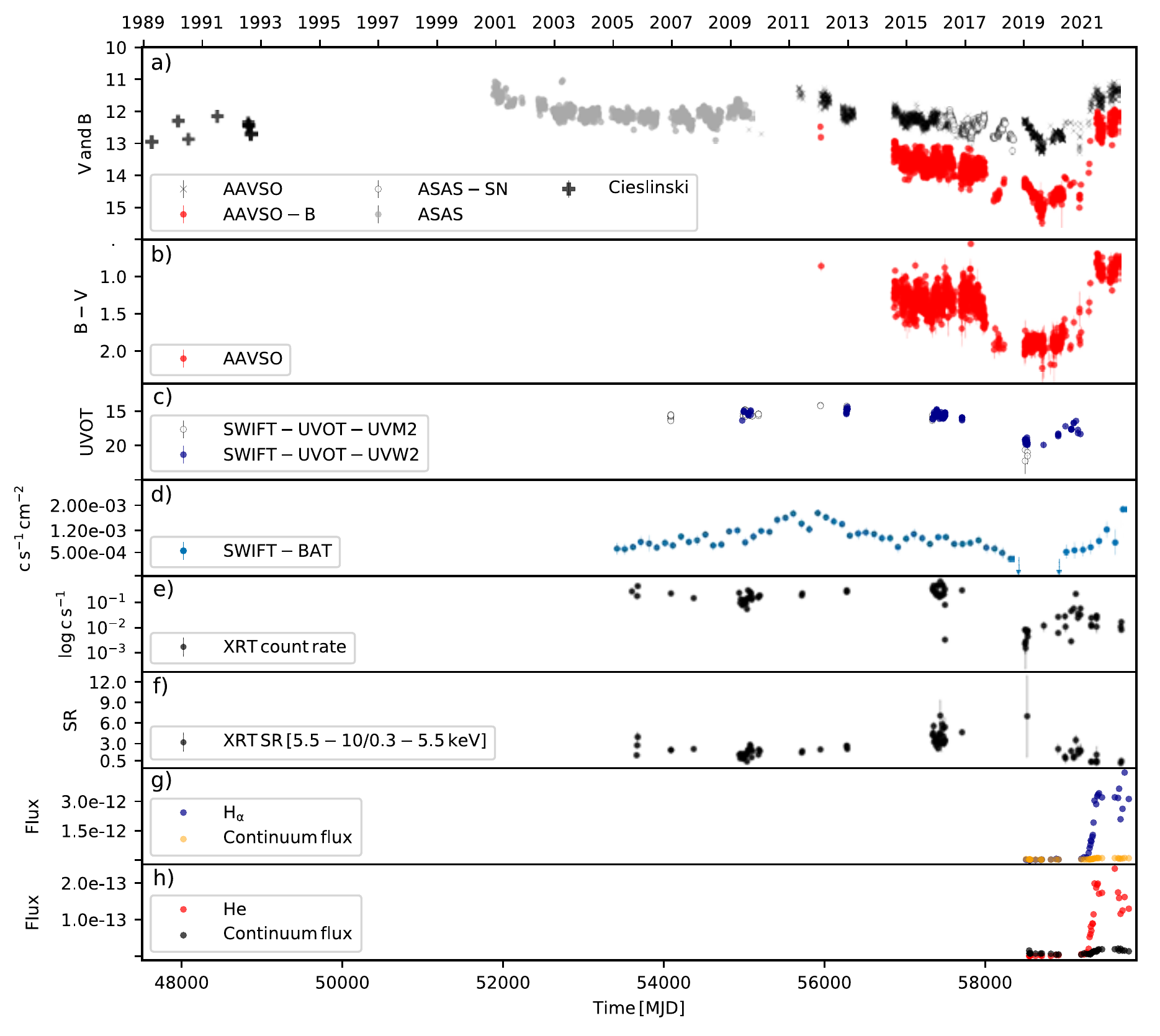}
\caption{Long-term optical to hard-X-ray variability of RT~Cru. {\it Panel a:}  RT~Cru optical {\it B}- and {\it V}-band light curves with measurements from \cite[][black plus signs]{1994A&AS..106..243C}, ASAS (gray dots), AAVSO (black crosses for V band, red circles for B band), and ASAS-SN (open circles). {\it Panel b:} B-V curve from AAVSO data. {\it Panel c:}
\swift/UVOT UVM2 (open circles) and UVW2 blue dots light curves. {\it Panel d:} Light blue dots are \swift/BAT 14--50 keV measurements with 100-day bins (with 1$\sigma$ error bars). The maximum in the BAT light curve around 2011-2013 is due to an increase in the accretion rate which did not increase enough to reach the level where the boundary layer becomes optically thick and a super-soft X-ray component arise \citep[see][]{2018A&A...619A..61L}. Between late 2019 and late 2020, RT~Cru was not detected by \swift/BAT in 100-days bins.{\it Panel e:} \swift/XRT 0.3-10 keV count rate in log scale to highlight the fade in flux during 2019. {\it Panel f:} \swift/XRT softness ratio. {\it Panel g:} H$\alpha$ (blue dots) and adjacent continuum (yellow dots) flux evolution. {\it Panel h:} He I 6678 \AA~ (red dots) and adjacent continuum (black dots) flux evolution. Starting around mid-2017, the B-V color became more red for a few
years (panel {\it b}) while the hard X-ray flux dipped (panel {\it d}). The UV and soft X-ray
emission (panels {\it c} and {\it e}) also dipped at around that same time.}
\label{fig1}
\end{figure*}

Figure \ref{fig1} shows AAVSO, ASAS, ASAS-SN optical photometry plus photometric observations extracted from \citet{1994A&AS..106..243C}, \swift/BAT/XRT/UVOT light curves and H$\alpha$ and He I 6678 \AA~ flux evolution. 

\subsection{\xmm  \label{sec:obsXMM}}

RT~Cru was observed with XMM-Newton for 58.8\,ks on 2019 March 3 following our request through Director's Discretionary Time (ObsID 0831790801). The EPIC MOS1, MOS2 and pn cameras were operated in Full Window mode, with the optical blocking medium filter. We selected standard grades for the PN (0--4) and MOS 1/2 (0--12) cameras. The RGS1 and RGS2 cameras were operated in Spectroscopy mode but rendered spectra with low signal-to-noise, useless for performing spectral analysis. The Optical Monitor in fast-mode provided photometry by switching between the UVW1, UVW2 and UVM2 filters.

Data were reduced with the Science Analysis Software (SAS) version 20.0.0. Removing intervals with high flaring background, the net exposure time of the EPIC cameras was reduced to 36.4 ks. X-ray spectra from the source and background were constructed using data extracted from circular regions with 20 and 45 arcsec radii, respectively. The source region was centered on RT~Cru's optical coordinates, while background was selected in a nearby, source-free area. We used the \texttt{rmfgen} and \texttt{arfgen} scripts to build the redistribution matrices and ancillary EPIC responses. The energy channels were grouped in such a way that each spectrum had at least 25 counts per bin. 
In the case of the OM, we used the \texttt{omfchain}/SAS task to extract photometric series from the UVW1 filter with a bin size of 120 s. RT~Cru was not detected in the UVW2 and UVM2 filters.

\subsection{\swift }

\swift/XRT and UVOT pointed observations aside from those already reported in \citet{2018A&A...616A..53L} were obtained through Target of Opportunity request every 1--2 months from 2019 January 15 to 2022 April 24. 

\subsubsection{\swift/BAT}

The \swift/BAT light curve in the 14--50 keV was extracted from the \swift/BAT Hard X-ray Transient Monitor \citep{bat_monitor} and binned at 100 days using the HEASOFT tool \texttt{lcurve} (see Panel {\em d} in Figure \ref{fig1}).

\subsubsection{\swift/XRT \label{sec:swfitXRT}}

\begin{figure}[ht!]
\includegraphics[scale=0.50]{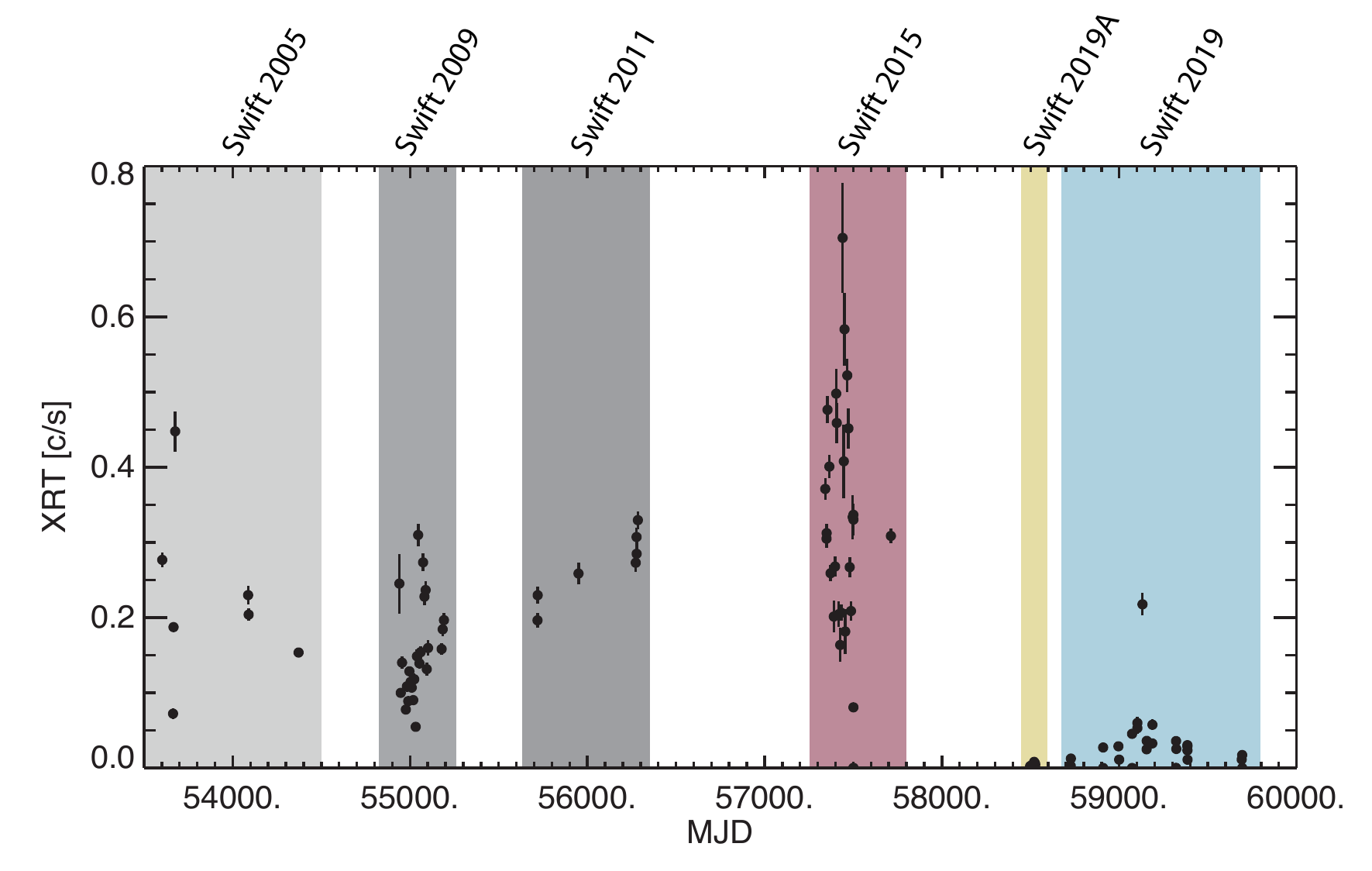}
\caption{\swift/XRT count rate light curve in the 0.3--10 keV energy range. The colored boxes show the observations that were combined to obtain the spectra analyzed in Section \ref{sec:swfitXRT}.}
\label{fig:xrtlc}
\end{figure}

Since 2005, \swift\ has observed RT~Cru with the XRT more than a hundred times, with a total exposure time of 240 ks. The exposure time of each of these snapshots was generally short, less than 2 ks on average, and thus the number of counts in each spectrum was low, precluding spectral modeling of the spectra individually. The XRT light curve from the individual pointings is clearly variable on the short and long terms (see panels {\em e} and {\em f} in Fig. \ref{fig1}). With the goal of studying the long term behavior in X-rays, we selected events with grades 0--12 and constructed six spectra from the XRT data by combining the event files from 2005 August 20 to 2007 September 29 ({\it Swift 2005}), 2009 April 19 to 2009 December 26 ({\it Swift 2009}), 2011 June 7 to 2012 December 12 ({\it Swift 2011}), 2015 November 16 to 2016 April 23 ({\it Swift 2015}), 2019 January 15 to 2019 February 12 ({\it Swift 2019A}) and 2019 September 1 to 2022 April 24 ({\it Swift post-2019}). Figure \ref{fig:xrtlc} shows the XRT light curve and the intervals from which the event files were combined and spectra extracted.

We extracted a long-term light curves and softness ratios (see Panels {\em e} and {\em f} in Figure \ref{fig1}) during each XRT pointing by selecting those photons enclosed in a 12 arcsec radius  circular region centered on coordinates provided by the tool \texttt{xrtcentroid}, which did not differ significantly from the optical coordinates. The corresponding background was extracted from an annulus region of 60 arcsec inner radius and 95 arcsec outer radius centered on the same coordinates as the source region. The same regions were used to extract the combined spectra, and the corresponding ancillary matrices were generated using the HEASOFT tool \texttt{mkrarf} while the response matrices were downloaded from the Calibration site of the HEASARC.

Very bright {\it optical} sources tend to generate artifacts that mimic X-ray photons or change the grades and energies of the X-ray photons, a process known as optical loading. Given that on average, the V optical magnitude of RT~Cru is 12 mag, we emphasize that both grade distribution and offset maps, as expected, indicate that the \swift/XRT and \xmm/EPIC data were not affected by optical loading, and that the soft X-rays in their spectra are real (see \url{https://www.swift.ac.uk/analysis/xrt/optical\_loading.php} for the optical loading thresholds in \swift/XRT).

\subsubsection{\swift\ UVOT}

The Ultra-violet Optical Telescope (UVOT) aboard \swift\ observed concurrent with the XRT instrument in either a lenticular filter or the UV grism.  The lenticular filters used were one of $UVW2$, $UVM2$, $UVW1$, $U$, or $V$.  The images were used to check that the observation was good, i.e., the source did not fall on one of the patches of lower sensitivity, and the pointing was stable. Calibrated magnitudes and the corresponding errors \citep{2008MNRAS.383..627P,2011AIPC.1358..373B} were obtained using the HEASoft V6.28 Ftool \texttt{uvotsource} which had been modified to flag data falling on a low-sensitivity patch on the detector\footnote{\url{https://heasarc.gsfc.nasa.gov/docs/heasarc/caldb/swift/docs/uvot/uvotcaldb\_sss\_01b.pdf}}.
The \swift/ UVOT CALDB version used was 20201019, which includes the most recent updated calibration files accounting for sensitivity changes\footnote{\url{https://heasarc.gsfc.nasa.gov/docs/heasarc/caldb/swift/docs/uvot/uvotcaldb\_throughput\_06.pdf}}.
The observations with the grism were checked for substantial contamination by any of the many sources in the crowded field.  The valid roll angles varied throughout the year, and data for some dates were not usable. Also, the UV brightness in RT~Cru was seen from time to time to drop below the detection level, in which case RT~Cru was only detected above 4,000\AA.  In order to obtain good S/N, extracted UVOT spectra from exposures spanning one day at most were summed. We used the calibration by \citet{2015MNRAS.449.2514K}, and the spectral extraction software from \citet{2014ascl.soft10004K}.

\subsection{Optical spectroscopy}

We obtained 31 spectra of RT~Cru between 2019 January 26 and 2022 April 15 (see observing log in Table \ref{tab:chironlog}) using the \textrm{Chiron} spectrograph \citep{2013PASP..125.1336T}. \textrm{Chiron} is a bench mounted fiber fed cross-dispersed echelle spectrograph on the 1.5m on the SMARTS telescope at the Cerro Tololo InterAmerican Observatory. We took data in ``fiber mode'', with 4$\times$4 on-chip binning yielding a resolution $\lambda/\delta\lambda \approx$27,800. Exposure times range from 10 to 30 minutes in 10 minute integrations.

The data were reduced using a pipeline coded in IDL (Walter 2017). The images were flat-fielded. Cosmic rays were removed using the L.A. Cosmic algorithm  \citep{2001PASP..113.1420V}. The 74 echelle orders were extracted using a boxcar extraction, and instrumental background, computed on both sides of the spectral trace, was subtracted. As $Chiron$ is fiber-fed, there is no simple method to subtract the sky. The fibers have a diameter of 2.7~arcsec on the sky. In any event, for bright targets like RT~Cru, night sky emission is generally negligible apart from narrow [OI] and Na D lines and some OH airglow lines at longer wavelengths. Wavelength calibration uses ThAr calibration lamp exposures at the start and end of the night, and occasionally throughout the night. \textrm{Chiron} in fiber mode is stable to better than 250 m s$^{-1}$ over the course of many nights.

The instrumental response was removed from the individual orders by dividing by the spectra of a flux-standard star, $\mu$~Col. This provides flux-calibrated orders with a systemic uncertainty due to sky conditions. The individual orders were spliced together, resulting in a calibrated spectrum from 4083-8900 \AA. Finally, we used contemporaneous $BVRI$ photometry to scale the spectrum to approximately true fluxes. 
\subsection{Optical photometry}

\subsubsection{ASAS and AAVSO}

For the ASAS data, we only considered those with quality flag "A" in the GRADE column and apertures with radii less than 30 arcsec. In the case of AAVSO, we selected V and B magnitude measurements, ignoring those data without measurement error bars. 

\subsubsection{{\it Swope}, {\it Du Pont} and {\it Gemini} telescopes}

In the framework of a long-term optical survey aimed at searching and characterizing rapid photometric variability (i.e., flickering) in southern symbiotic stars and nova-like objects \citep{2012ApJ...756L..21A,2013IAUS..290..179A}, we have been monitoring RT~Cru since 2012 with various facilities in Chile. The flickering data presented in this paper were collected at the 1 m {\it Swope} telescope and at the 2.5 m {\it Du Pont} telescope of the Las Campanas Observatory 
as well as at the 8.1 m Gemini South telescope. 

Table \ref{tab-flickering} presents a list of the observations where, for each date, we report the MJD, calendar date, the maximum minus minimum magnitudes ({\em max-min}), average magnitude, band, facility (telescope/instrument combination), the overall length of each light curve and the duration of individual exposures.
For technical details about the telescope/instrument configuration (e.g., field-of-view, pixel-scale, read-out times, etc.), we refer the interested reader to the official observatory web sites  \footnote{\url{https://obs.carnegiescience.edu/swope}} \footnote{\url{https://obs.carnegiescience.edu/dupont}} \footnote{\url{https://www.gemini.edu
}}. The data reduction for this quite heterogeneous dataset was self-consistently performed with \texttt{THELI} v3.0.5 \citep{2013ApJS..209...21S}, a powerful and versatile tool for the automated reduction of astronomical images.  The differential ensemble photometry \citep{1992PASP..104..435H} was then performed with \texttt{VAPHOT} \citep{2013ascl.soft09002D, 2001phot.work...85D}, an aperture photometry package running within the \texttt{IRAF} environment, specifically developed for dealing with precise time-series photometry of uncrowded fields. Its characteristic feature is the ability to work with aperture sizes that have been finely tuned (through preliminary PSF fitting) to generate the best signal-to-noise ratio for each target/comparison star(s) in a single CCD frame. 

Because of RT~Cru's large brightness variation in the course of the last fifteen years, it was not possible to use the same exposure times nor the same set of comparison stars across the different observing runs. Moreover, even though the flickering survey was preferentially conducted through the Johnson U filter, during RT~Cru's low-state we had to resort to the B filter in order to both keep the sampling rate of the flickering light curve reasonably short (i.e., under about 5 minutes) and, at the same time, reach suitable S/N values. We conservatively assumed as photometric error of our RT~Cru flickering light curve the standard deviation of the differential light curves built with the comparison stars used for the ensemble photometry (refer to \citet{2012ApJ...756L..21A} and \citet{2011ASSL..373...33M} for more details about this technique).

\subsubsection{Helen Sawyer Hogg (HSH) telescope}

Photometric observations in $B$, $V$ and $R$ bands were obtained with the 0.6m HSH telescope located at Complejo Astron\'omico El Leoncito, Argentina\footnote{Based on data obtained at Complejo Astronómico El Leoncito, operated under agreement between the Consejo Nacional de Investigaciones Científicas y Técnicas de la República Argentina and the National Universities of La Plata, Córdoba and San Juan} from 2021 February 15 to 2022 May 4 (see also Table \ref{tab-flickering}). The telescope is equipped with a SBIG STL-1001E CCD camera and $UBVRI$ Johnson-Cousins filters. 

The data reduction and aperture photometry were performed by using \texttt{AstroImageJ} \citep{2015AAS...22533616H}. The data reduction included: bias and dark subtraction and flat field correction. The photometric standards from the APASS catalogue \citep{astroimagej}, 2MASS 12345405-6435040 (T2); 2MASS 12345752-6435099 (T3) and 2MASS 12345855-6436071 (T4) were used to obtain absolute magnitude of RT~Cru. These stars are located at an angular distance of less than 2.2 arcmin from RT~Cru. The observing log is listed in Table~\ref{tab-flickering} while light curves from each observing run are displayed in Fig. \ref{fig.rtcru.hsh}.

\subsubsection{TESS}

The Transiting Exoplanet Survey Satellite \citep[TESS;][]{ricker14} observed the field of RT Cru during sector 11 (2019 April 22 - May 21) and again in sectors 37 and 38 (2021 April 2 - May 26). Cadences were 30 minutes in sector 11, and 10 minutes in sectors 37 and 38. For all sectors, the photometric accuracy was 0.001 mag. TESS is a single-channel photometer with a 6000--10000~\AA\ bandpass.

We downloaded 40x40 pixel cuts from the full frame image data from the MAST archives using the TESScut software \citep{brasseur19}. TESS images, while photometrically stable and of continuous cadence, suffer from coarse spatial resolution  (21\arcsec\ pixels). We extracted the data using aperture photometry with a 2.5 pixel radius. Background is extracted from an annulus between 5 and 10 pixels from the source. Because there are often other sources in the background annulus, we iteratively select the background pixels, removing those more than 3$\sigma$ from the median level until we converge on the median background level. We assume that the background is spatially flat in this region.

\begin{figure*}
\begin{center}
\includegraphics[scale=0.7]{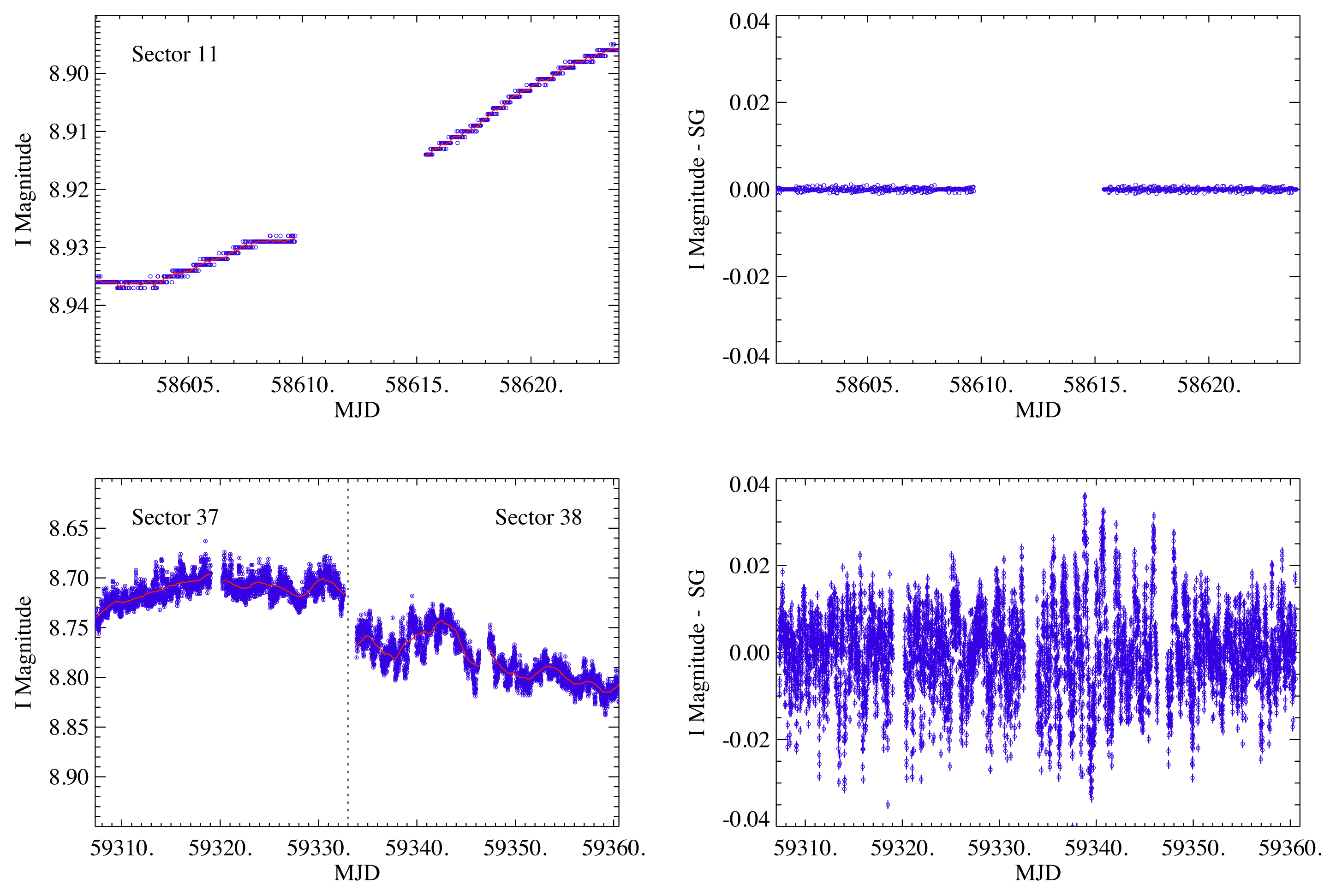}
\caption{{\it TESS} light curves of the observations obtained during Sector 11 in 2019 with a 30 m cadence (top row), and the light curves from observations obtained during Sectors 37 and 38 in 2021 with a 10 m cadence (bottom row). The vertical dashed line in the lower left-hand plot marks the separation between Sectors 37 and 38. Right panels: Each portion of the light curve of each sector has a Savitzky-Golay filter (red line in the left panels) subtracted to enable the study of short-term, flickering-type variability. Right-hand panels show the light curves after the SG filter was subtracted. Variability on time scales shorter than 10 min (in Sectors 37 and 38) or 30 min (in Sector 11) cannot be detected.
In addition, the measurement errors (0.001 mag) are larger than the observed dispersion; no short-term variability can be detected. }
\label{lc:tess}
\end{center}
\end{figure*}

Figure \ref{lc:tess} shows the light curves of Sectors 11, 37 and 38. There is a gap in each sector because the spacecraft is re-oriented to dump data in the midst of each sector. Consequently, we 
analyzed each half of the $~$27 day light curve of each sector separately. To study the short 
(minutes to a few hours) time scales, we subtracted from each light curve a Savitzky-Golay \citep[SG;][]{1964AnaCh..36.1627S} filter to remove the long-term (days) trend. 

We searched for periods in the light curve by using $R$\footnote{\url{https://www.r-project.org
}} implementation of the REDFIT algorithm \citep{2002CG.....28..421S} which uses the Lomb-Scargle algorithm, accounts for and models the red noise and performs Monte Carlo tests for computing red noise false-alarm levels at 1, 2 and 3$\sigma$ levels. 

\subsubsection{Classification of the optical spectrum during the optical minimum.}

In contrast to the optical spectra during 2018 presented in \citep{2018A&A...616A..53L}, the optical spectra taken during 2019 reminds one of a single red giant. We first corrected the optical spectrum obtained on 2019 March 17 with SMARTS and the January 2019 UV spectrum obtained with \swift/UVOT for interstellar reddening using $E_{B-V}$= 0.53 (roughly corresponding to the interstellar absorption toward RT~Cru as provided by the STructuring by Inversion the Local Interstellar Medium team \footnote{\url{https://stilism.obspm.fr/}} \citep{2017A&A...606A..65C}) and the extinction law of \citet{1989ApJ...345..245C}. We then performed spectral classification of March SMARTS spectrum with the Python package PyHammer \citep{2017ApJS..230...16K}, which enables comparison of various spectral templates to the spectrum under investigation, and visual inspection. PyHammer gives spectral type M5 as the best fit to the spectrum. The precision with which PyHammer works is approximately one spectral subtype \citep{2017ApJS..230...16K}. The spectrum was normalized by the flux at 8150 \AA. Using the minimum V-band magnitude in 2019 when the optical emission lines disappeared ($m_v=13.303\pm0.026$, from the AAVSO database) and the distance of $2.52\pm0.19$ kpc, we calculated $M_v=-0.28\pm0.18$. We find the luminosity class of RT~Cru to be III, as derived from spectral type-absolute magnitude ($M_V$) calibration (Table II in  \citet{1981Ap&SS..80..353S}). The B-V color from early 2017 through late 2020 indicates that the emission from the system became redder during that time (see Fig. \ref{fig1} panel {\it b}), supporting our contention that the red giant emission overwhelmed that from the WD and/or the accretion disk. Comparing the spectrum with templates of M giants by \citet{2015RAA....15.1154Z} we find that the spectrum was closer to the M6III template. In Figure \ref{fig:RTCru_specclass} we plot the 2019 March 17 spectrum and the UVOT January 2019 with the template obtained from \citet{2015RAA....15.1154Z}. The UVOT spectrum is also compatible with that of a single M6 III red giant (the flux calibration above 4,000\AA~ is highly uncertain due to possible contamination of higher order spectra and was removed from the data).

\section{Results \label{sec:res}}

Below, we describe evidence that the changes in the observational appearance of RT~Cru beginning in  mid-2017 -- namely the vanishing of characteristic symbiotic-star optical emission features along with the scaling down of the flickering amplitude and the X-ray luminosity -- were the result of a strong drop in the rate of accretion onto the WD. This finding was possible thanks to the  broad-band energy coverage, which allowed us to measure the accretion rate via modeling of the X-ray spectrum, along with quasi-simultaneous observations of phenomenological changes at other wavelengths.

\subsection{The disappearance of optical emission lines in 2019}

\begin{figure}
\includegraphics[scale=0.35]{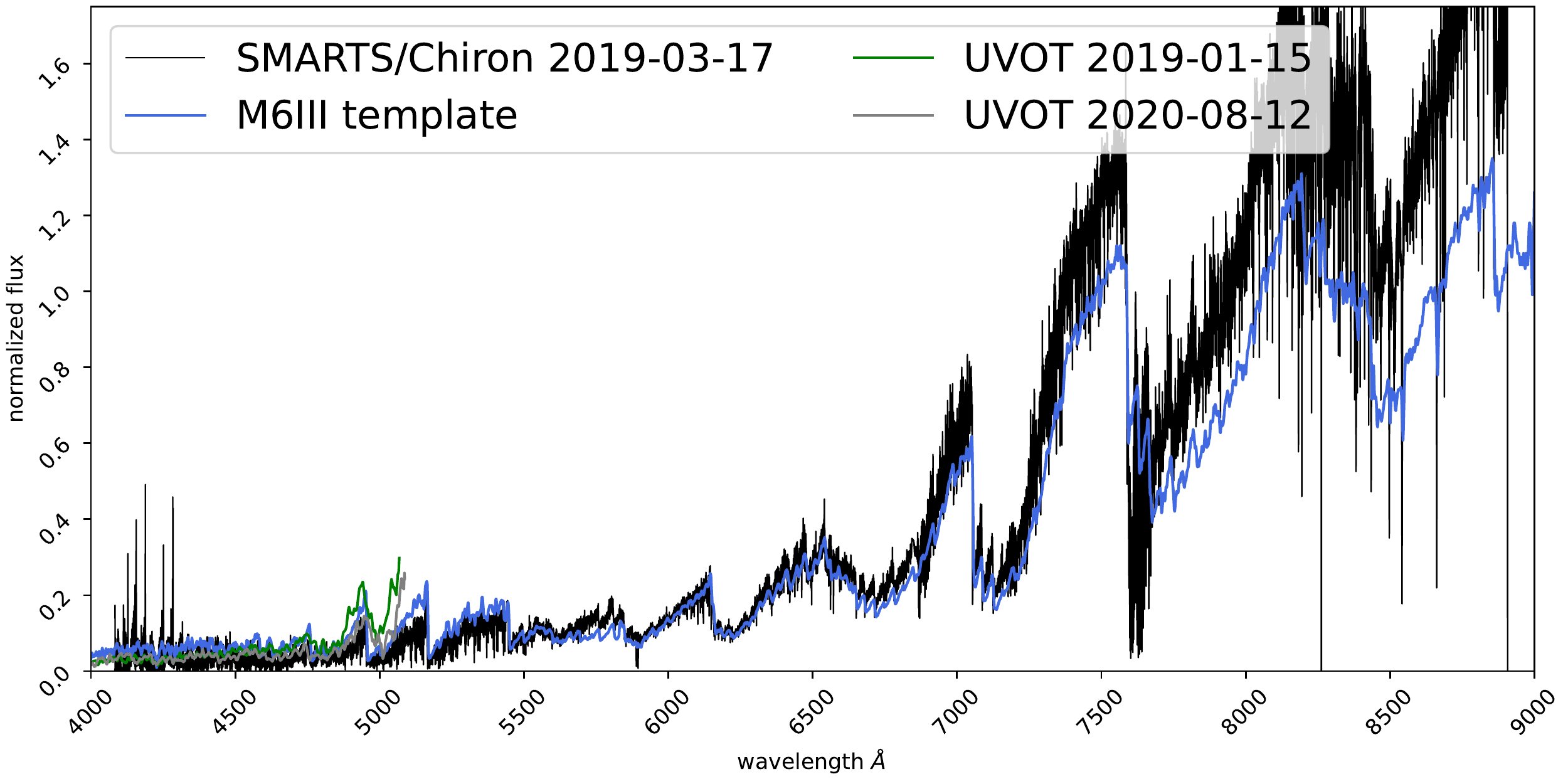}
\caption{The SMARTS optical spectrum of RT~Cru obtained on 2019-03-17 
(black line) and the UVOT spectra obtained on 2019-01-15 (green line) and 2020-08-12 (gray line),
corrected for the interstellar extinction using E(B-V)=0.53. The blue line is a template spectrum of a M6III giant from \citet{2015RAA....15.1154Z}.} 
\label{fig:RTCru_specclass}
\end{figure}

The first hint of a decrease in the accretion rate came from the disappearance of the typically-observed Balmer emission lines, with the optical spectrum coming to resemble that of a single red giant (see Fig. \ref{fig:RTCru_specclass}).

\begin{figure*}
\includegraphics[scale=0.7]{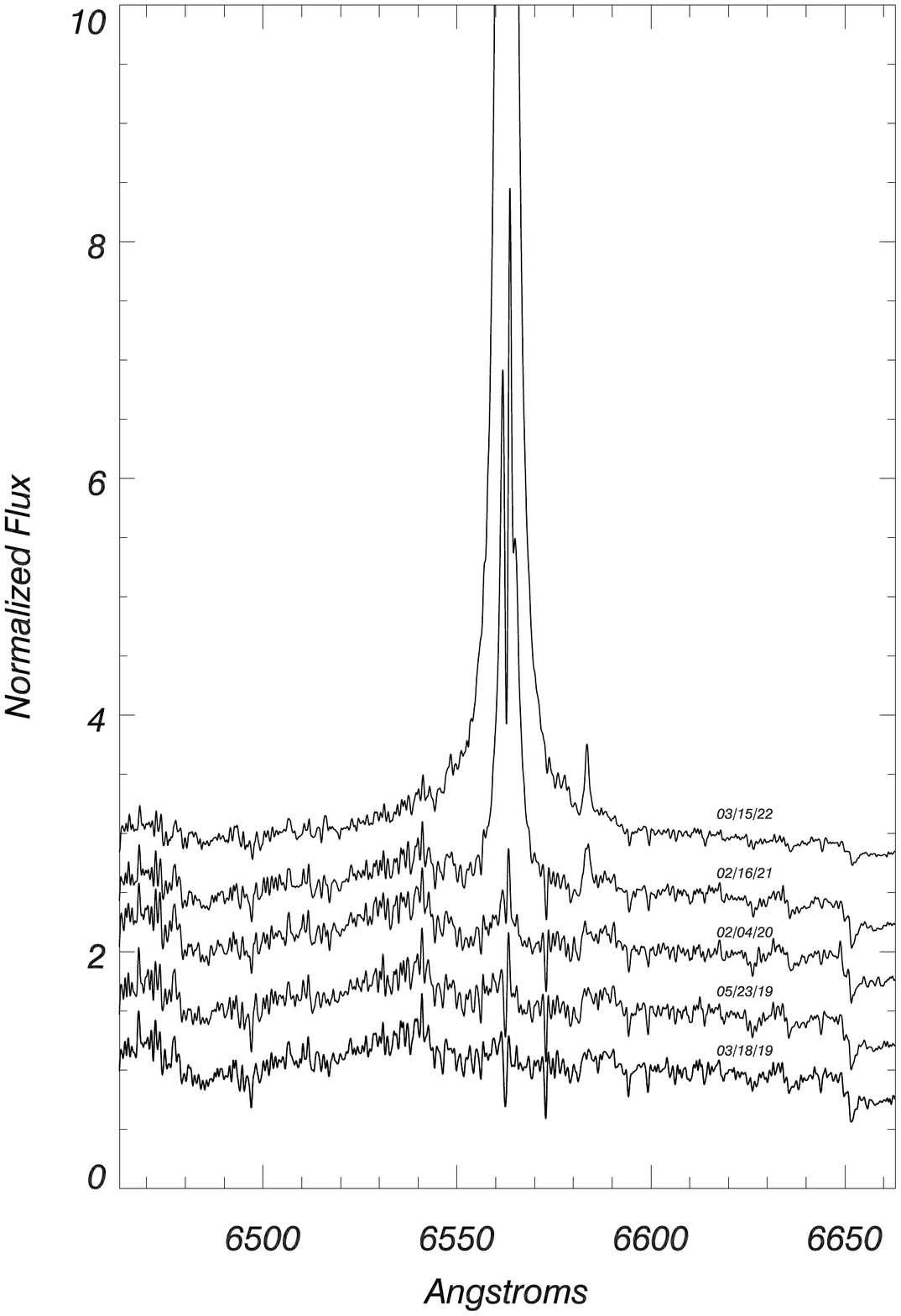}
\caption{Selected optical spectra taken with the SMARTS/Chiron telescope from March 2019 through March 2022 (see the complete list of spectra in Table \ref{tab:chironlog}). Since March 2019, H$\alpha$ is detected in emission, while its flux increased monotonically (see Panel {\em g} in Fig. \ref{fig1}). 
Other lines such as N[II] 6584 \AA~ were also missing in 2019 and reappeared in emission afterwards.}
\label{fig:opt_spec}
\end{figure*}

In Figure \ref{fig:opt_spec} we show a series of optical spectra taken from March 2019 to April 2022, when it can be seen that H$\alpha$ \citep[usually the strongest emission line in RT~Cru as observed in spectra taken in 2012; see][]{2018A&A...616A..53L} 
was in absorption 
through March 18, and was weakly in emission by May 22 2019, after which the strength grew more-or-less monotonically (see panel {\em g} in Fig. \ref{fig1}). 




\subsection{The reduction of rapid optical and UV variability in 2019 \label{timing-aper}}

\begin{figure*}
\begin{center}
\includegraphics[scale=0.6]{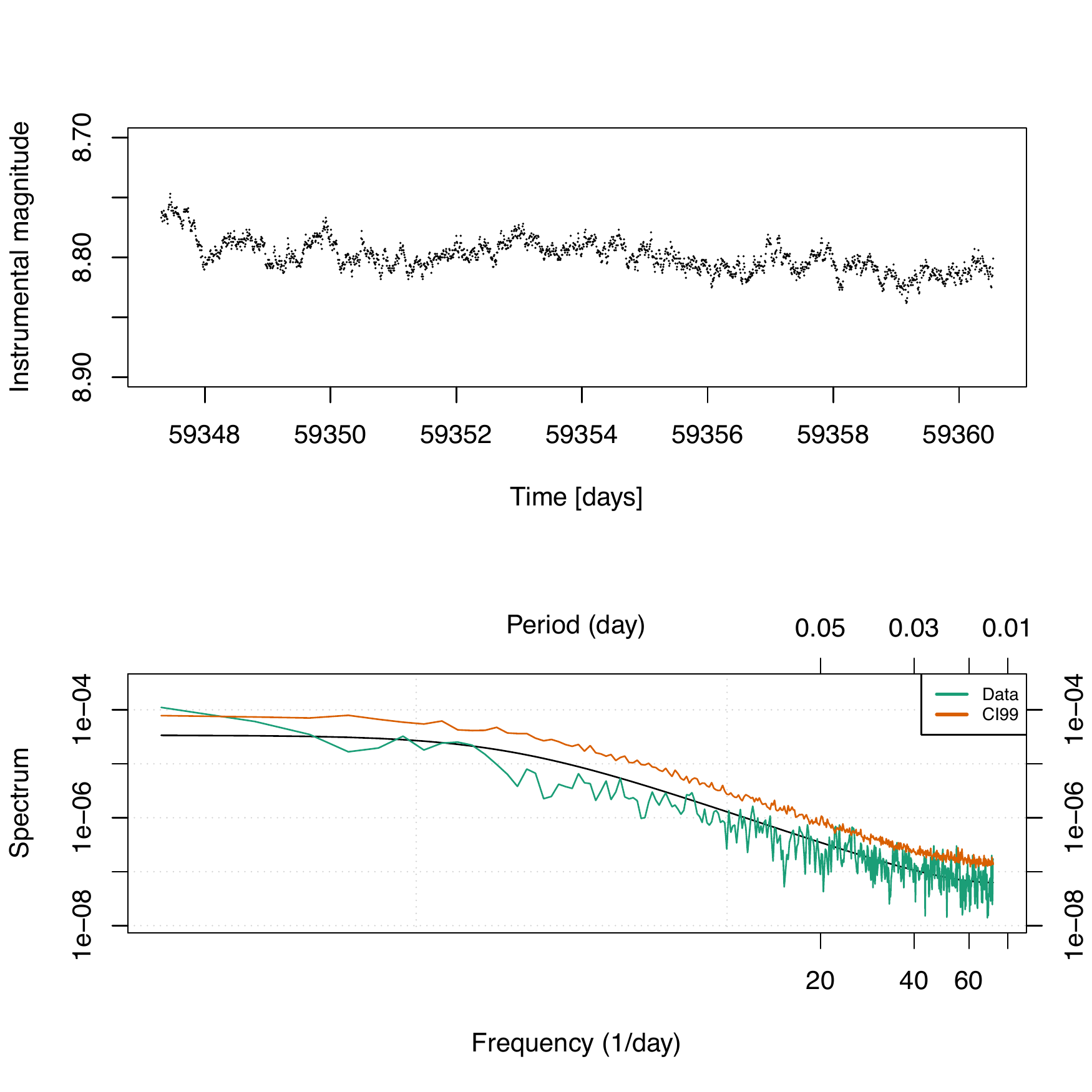}
\caption{{\em Top}: {\it TESS} light curve of the second half of Sector 38. {\em Bottom}: LS power spectrum (green line) with red noise model (black line) and detection levels at 3$\sigma$ (orange).}
\label{lc:tess38a}
\end{center}
\end{figure*}

\begin{figure}[ht!]
\includegraphics[scale=0.5]{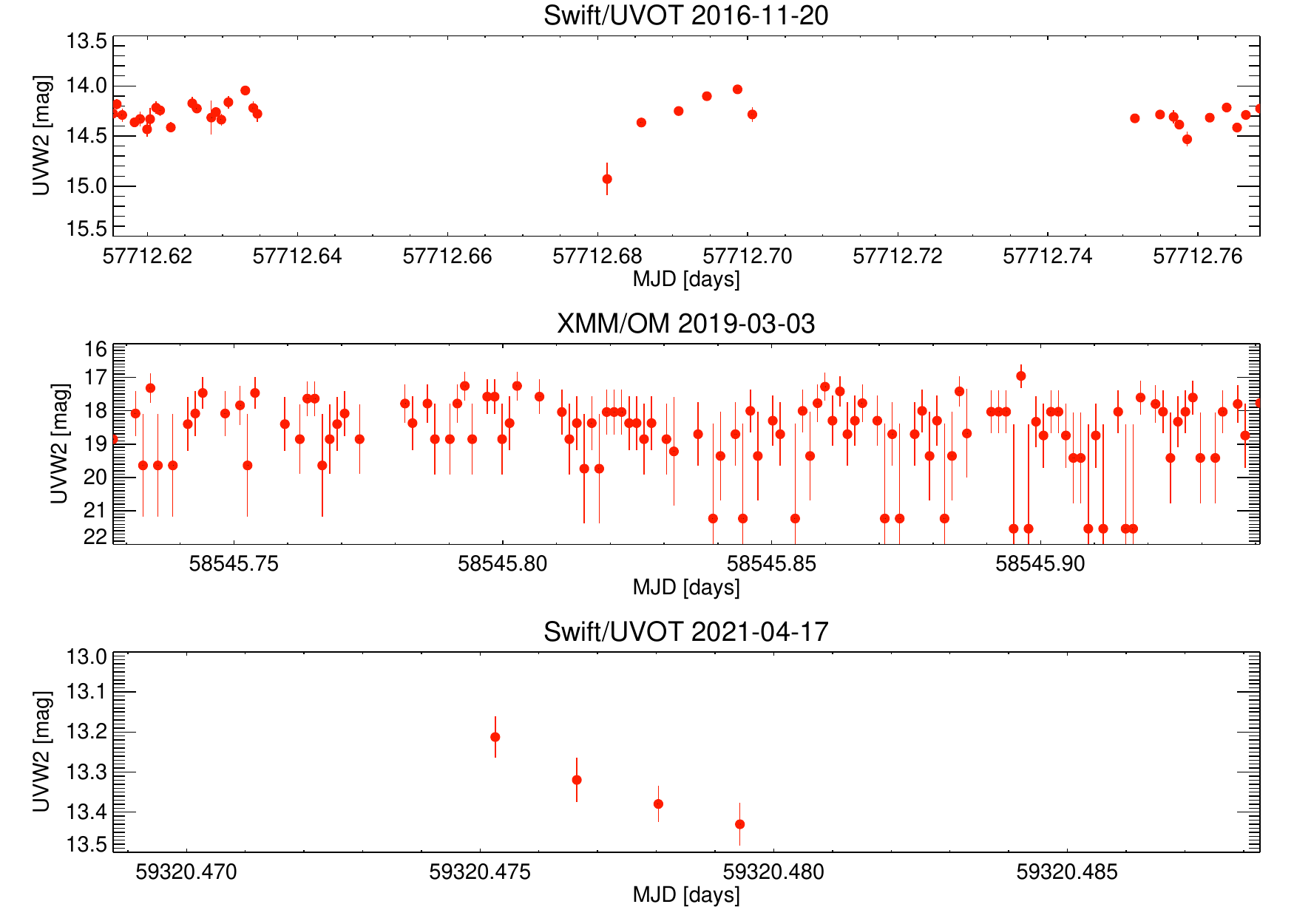}
\caption{UVW2 \swift/UVOT and \xmm/OM light curves with 120 s bins. The UVOT light curves from dates outside the low-flux optical state (years 2016 and 2021; top and bottom panels) show strong variability whereas the light curve from the 2019 \xmm/OM observation (middle panel) shows a much fainter source with no significant variability.}
\label{fig:om}
\end{figure}

In symbiotics powered by accretion alone, a hallmark of the presence of the accretion disk is the strong variability in optical/UV brightness observed on time scales of a few seconds to a few minutes \citep{2001MNRAS.326..553S, 2013A&A...559A...6L}, namely,  flickering. Since its discovery, RT~Cru has shown strong flickering in every wavelength observed, from X-rays to optical \citep{1994A&A...284..156S, 2016A&A...592A..58D, 2018A&A...616A..53L,2021MNRAS.500.4801D}. 
Our ground-based optical photometric observation runs spanned on average a few hours, with varying exposure times during different observing runs. In order to study the flickering amplitude in AAVSO data, we grouped those observation taken within a single night and with more than 70 measurements with a maximum cadence of 1 min. The {\it TESS} light curves show significant contribution of red noise during Sectors 37 and 38, whereas during Sector 11 (observed in April 2019), the light curve did not show significant variability on short time scales. No periods were detected above a detection threshold of  1, 2 and 3$\sigma$ (see Fig. \ref{lc:tess38a}). The flickering amplitudes from the ground-based observations and $TESS$, measured as 
{\em max-min} in each light curve, are listed in Table \ref{tab-flickering}. During most of 2019 the amplitude clearly decreased, as evidenced by the small amplitude of flickering in the $U$ and $B$ light curves in 2019 February taken with the {\it DuPont} telescope, in 2019 March and July taken with the {\it Swope} telescope and in 2019 April and May taken with $TESS$ (see Figures \ref{fig.rtcru.dupont}, \ref{fig.rtcru.swope} and \ref{lc:tess}). The $UVM1$ \xmm/Optical Monitor light curves obtained during the 2019 observation show a much fainter UV emission than during \swift\ observations in 2016 and 2021, with high photometric errors, thus the uncertainty in the amplitude is comparable to the amplitude itself ({\em max-min}=0.76$\pm$1.07), pointing to unusually small variability consistent with the major drop in the amplitude of the rapid variability seen from the ground (see Figure \ref{fig:om}). 

\subsection{The decrease in absorbing column and accretion rate from the X-ray spectrum \label{sec:xfit}}

By modeling the X-ray spectrum, we can estimate the accretion rate through the optically thin emitting boundary layer \citep[e.g.,][]{2018A&A...619A..61L}. The hard X-ray spectrum of accreting WDs can be described by that of an isobaric cooling flow, where the plasma is radiatively cooling from a high post shock temperature under constant pressure. Given the high temperature of the X-ray emission, the most likely region where this emission arises is the accretion disk's boundary layer, where strong enough shocks are expected in the case of a massive white dwarf. In such case, its luminosity should be half of that provided by the accretion flow. The \texttt{mkcflow} XSPEC model yields the accretion rate through its normalization. We fit the \swift-combined spectra (see Section \ref{sec:swfitXRT}) with a model similar to that used to model earlier X-ray spectra from \nustar\ (obtained in 2016 November 22) and \suzaku\ (obtained in 2012 February 06) in \citet{2018A&A...616A..53L}, i.e. full-covering and partial covering absorbers that modify an isobaric cooling flow plus a Gaussian profile to account for the presence of the Fe K$\alpha$ fluorescence line at 6.4 keV (\texttt{TBabs$\times$pcfabs$\times$(mkcflow+gauss))}. In the spectra \swift 2005, \swift 2009, \swift 2011 and \swift 2015 an additional optically thin thermal model (\texttt{apec}) is needed to fit the emission at energies below 1 keV. This extra emission is consistent with $\beta$-type emission \citep{1997A&A...319..201M} and as it is not subject to the same absorption than the cooling flow component, we suggests that is arise in a colliding-winds region. We note, however, that due to the low number of counts in this spectral region, the resulting parameters from the fit can not be constrained. The \xmm\ (2019 March) and \swift/2019 spectra (2019 January -- February) required a simpler model without the need of a partial covering absorber or an additional thermal component. 

\begin{figure*}
\begin{center}
\includegraphics[scale=0.7]{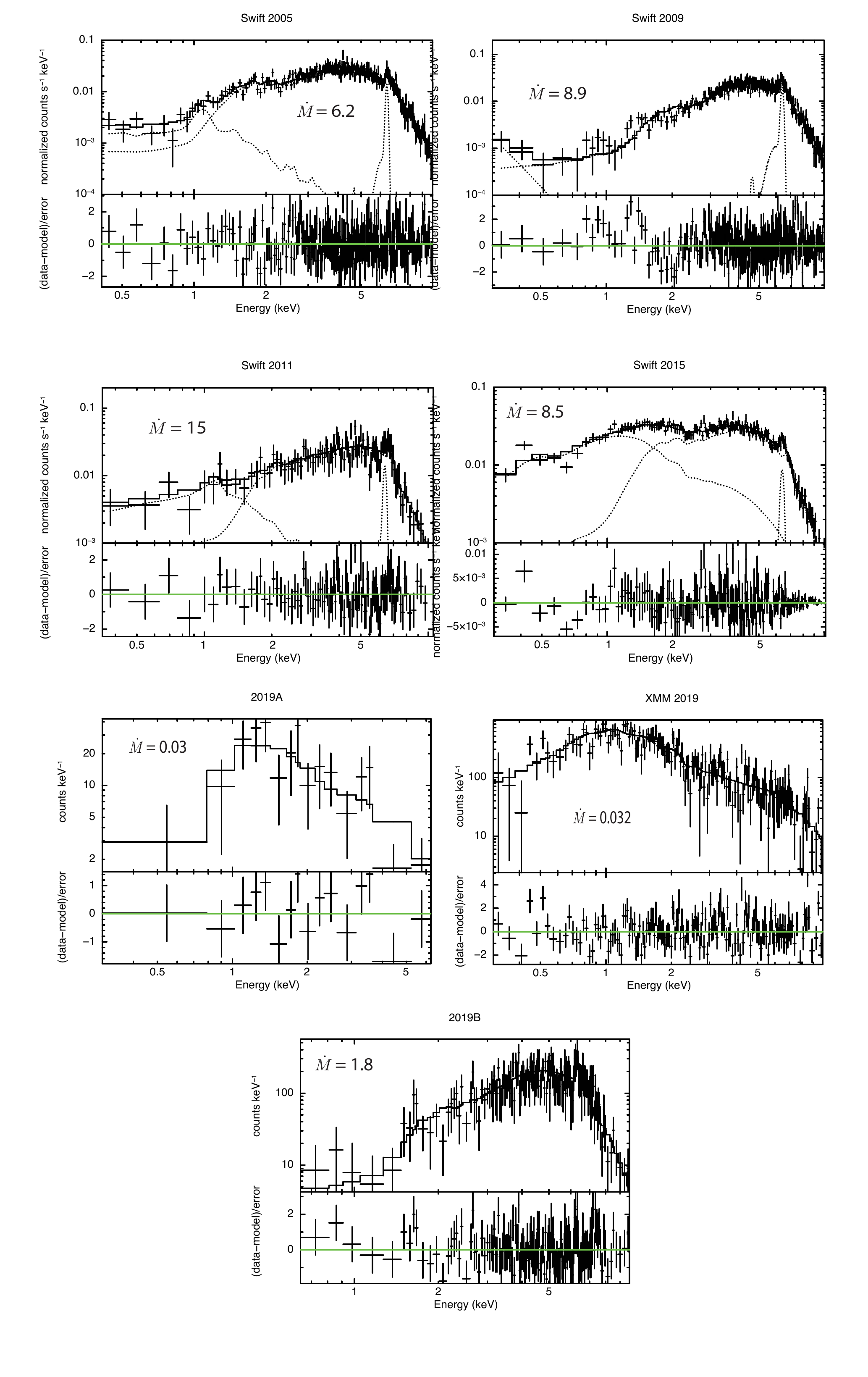}
\caption{X-ray spectral evolution between 2005 and 2022, as revealed by \xmm~ and \swift. The solid curves in the top portion of each panel are the best-fit models, while dotted lines shows the contribution of each spectral component in the cases where more than one spectral component was needed to model the spectrum (see Section \ref{sec:xfit}). Each panel lists the accretion rate through the optically thin boundary layer ($\dot{M}$, in units of 10$^{-9}$ \ms yr$^{-1}$). The resulting fit parameters are listed in Table \ref{tab:xray}. The spectra taken in 2019 with \swift\ and \xmm\ clearly show a decrease both in absorption and accretion rate.} 
\label{fig_all_x}
\end{center}
\end{figure*}

The investigation of all spectra assumed the maximum temperature of 53 keV as determined in \citet{2018A&A...616A..53L}, allowing to vary freely the normalization of the \texttt{mkcflow} component, ultimately corresponding to the expected accretion rate through the optically thin portion of the accretion disk's boundary layer. The resulting parameters from the fit are listed in Table \ref{tab:xray}. Figure \ref{fig_all_x} shows the spectra and corresponding best-fit models, where we omitted the MOS 1 and MOS 2 \xmm\ spectra for clarity, but used them during the spectral modeling. The absorption column shows a significant decrease during the 2019 observations, with the lowest value during the \xmm\ observation on March 2019. In the course of low accretion periods, less material is available to absorb the X-ray emission from the inner accretion disk, and thus a lower absorption column results from the spectral modeling. If the orbital-modulated accretion rate scenario proposed by \citep{2018A&A...616A..53L} is correct, then the low accretion rate, low absorption periods take place during apastron. 

The accretion rates through the optically thin portion of the accretion disk boundary layer derived from the fit of X-ray spectra taken at the different epochs described in Section \ref{sec:swfitXRT} (see Table \ref{tab:xray}) show that the accretion rates during the \xmm\ and \swift~2019A spectra were the lowest measured in RT~Cru.

The \nustar~ spectrum analyzed in \citep{2018A&A...616A..53L} shows that the X-ray emission in the 14--50 keV energy range arise from the same cooling flow as the lower energy X-ray emission. Thus the \swift/BAT flux can be used to trace the accretion rate through the boundary layer. The \swift/BAT light curve indicates a low accretion rate period during a large fraction of 2019, when the flux reached a  minimum of 1.7 $\times$ 10$^{-5}$ cts cm$^{-2}$ s$^{-1}$. For comparison, during the maximum in 2011-2013, the BAT flux reached 1.8 $\times$ 10$^{-3}$ cts cm$^{-2}$ s$^{-1}$ and outside the maximum or the minimum, it has an average flux of 9$\times$ 10$^{-4}$ cts cm$^{-2}$ s$^{-1}$.

\begin{sidewaystable*}
\caption{X-ray spectral model parameters (see Section \ref{sec:xfit}). Unabsorbed luminosity in the 0.3-10 keV energy range at a distance of 2.52 kpc. The parameters of the cooling flow component are indicated with $CF$.}
\label{tab:xray}
\begin{tabular}{ccccccccccc}
\hline
Date range & Instrument & kT & n$_H$ &L$_{X}$ & n$_H^{CF}$ & n$_H(PC)^{CF}$ & CF & $\dot{M}^{CF}$ & L$_{X}^{CF}$ \\
 &  & [keV] &[10$^{22}$ cm$^{-2}$] & [10$^{31}$\lumcgs]& [10$^{22}$ cm$^{-2}$] & [10$^{22}$ cm$^{-2}$] & & [10$^{-9}$ \ms~yr$^{-1}$] & [10$^{31}$\lumcgs] \\
 \hline\hline
2005-08-20/2007-09-28 &	\swift & 1.9$_{-0.5}^{+4}$ & $\lesssim$0.3 & 9.3 &3.5$\pm$0.5	&20$\pm$3 &0.80$\pm$0.05& 6.2$\pm$0.4 & 4200\\
2009-04-19/2009-12-26 &	\swift & $\lesssim$3.2 & $\lesssim$0.3 &0.8 &6.7$\pm$0.5	& 35$\pm$5 & 0.86$\pm$0.03 & 8.9$\pm$0.6 & 5800\\
2011-06-07/2012-12-24 &	\swift & $\lesssim$1.6& $\lesssim$0.2 & 32 &6$\pm$2	& 65$\pm$25 & 0.85$\pm$0.04 & 15$\pm$4 & 10700 \\
2015-11-16/2016-11-20 &	\swift & $\lesssim$5 & 0.07$\pm$0.03 & 320 &3.8$\pm$0.8	& 33$\pm$5 & 0.73$\pm$0.04 & 8.5$\pm$1 & 5900\\
2019-01-15/2019-02-12  & \swift & \nodata&\nodata &\nodata&0.4$\pm$0.2 & \nodata & \nodata & 0.03$\pm$0.01 & 24\\ 
2019-03-03 & {\it XMM} & \nodata& \nodata&\nodata& 0.24$\pm$0.03 & \nodata & \nodata & 0.032$\pm$0.006 & 22\\
2019-09-01/2022-04-24 & \swift & \nodata&\nodata&\nodata & 4$\pm$1 & 27$\pm$7 & 0.88$\pm$0.06 & 1.8$\pm$0.3 & 1200\\ 

\hline
\end{tabular}

\end{sidewaystable*}

\section{Discussion and conclusions} \label{sec:disc}

We have witnessed a significant decrease in the flow of matter through the accretion disk in a symbiotic star and 
the subsequent recovery. This conclusion is supported by our observational data as follows:

\begin{itemize}
    \item The significant decrease in hard X-ray flux in the \swift/BAT band during most of the year 2019 in addition to the softness of 
the X-ray spectrum obtained with \xmm\ indicates that accretion through the optically thin portion of the boundary layer decreased by two orders of magnitude, down to 3.2 $\pm$ 0.06 $\times$ 10$^{-11}$ M$_{\odot}$ yr$^{-1}$ (d/2.52 kpc)$^{2}$, while the strong absorption of X-rays usually observed during high accretion rate periods decreased as well.

    \item The conclusion about the decrease in the accretion rate is supported also by our analysis of the rapid variability in the optical light curves. During 2019, the short term $U$, $B$, $V$, $UVW2$ and $TESS$ optical light curves did not show the rapid variability typically observed during high accretion, optically bright states. It is noteworthy that a significant decrease in $\dot{M}$ manifests itself only as a modest overall dimming in optical -- by 
    approximately 1 magnitude (see Fig. \ref{fig1}); it was enough however to cause a significant decrease in the flickering amplitude. Given that flickering is a hallmark of an accretion disk, its low amplitude indicates that accretion significantly decreased    during 2019 and recovered afterwards. 
    Decreases of the optical flickering amplitude have been observed in other symbiotics (e.g. MWC~560 \cite{2011IBVS.5995....1Z}, CH~Cyg \cite{2003ApJ...584.1027S}). Low/high X-ray states have been associated with low/high optical line flux states in SU~Lyn \citep{2016MNRAS.461L...1M,2018ApJ...864...46L,2022MNRAS.510.2707I}.
    We are not aware of other observations of symbiotic stars revealing such extreme multi-wavelength changes associated with a change in optical flickering.
    
    \item Observed changes in UV flickering also support our conclusion about reduced accretion onto the WD. Non-burning symbiotic stars with hard X-ray emission and X-ray luminosities of about 10$^{32}$ \lumcgs\ (in the 0.3-10 keV range) show detectable flickering in the \swift/UVOT photometric data, whereas those with lower X-ray luminosities have lower UV flickering amplitudes \citep{2013A&A...559A...6L}. During the low-state of RT~Cru reported here, the same trend is clear.
    
    
    \item The decrease in $\dot{M}$ also had significant effects on the optical spectra. The usually-strong optical emission lines such as H$\alpha$ and He I, were absent in the spectra taken during the beginning of 2019, indicating the luminosity of the ionizing source decreased as a consequence of the lower accretion rate. 
    
    
\end{itemize}    
    
The detection of flickering within the $TESS$ bandpass (6,000-10,000 \AA) indicates that the accretion disk contributes significantly to the SED above 6,000 \AA\ (see Figure \ref{lc:tess}). This might be the case in other symbiotics observed with $TESS$ (results of $TESS$ observations of symbiotics will be presented in a forthcoming article).

    
The transition in and out of the low-state, as observed at all wavelengths, took about 2--3 years. Other symbiotics with similarly massive white dwarfs but shorter orbital periods, such as T~CrB, have longer-lasting high and low states. The red giant in T~CrB is thought to fill its Roche-lobe \citep{1958ApJ...127..625K}, and thus the accretion disk is probably usually present and the brightening episodes such as the one reported in \citet{2018A&A...619A..61L,2020ApJ...902L..14L} are most likely the result of disk instabilities such as those in dwarf novae. In long-orbital-period symbiotics where accretion proceeds through wind-capture, as we suspect is the case in RT~Cru and SU~Lyn just to mention a few examples, multiple factors could control the accretion rate: 1) an eccentric orbit; 2) wind accretion leading to a transient disk \citep{2021MNRAS.501..201H}; 3) changes in the mass loss rate from the red giant donor \citep{2021MNRAS.501..201H}; 4) disruption of the inner disk after a collimated outflow is launched, as in the long-orbital-period symbiotic CH Cyg \citep{2003ApJ...584.1021S}; or
5) dwarf-nova-like disk instabilities \citep{1986A&A...163...61D}.


Given the estimated mass of the WD in RT~Cru of at least about 1.2\ms~ \citep{2007ApJ...671..741L}, it might seem somewhat surprising that no nova-type outbursts have been recorded during the last century. However, theoretical models by \citet{2005ApJ...623..398Y} predict that at a constant accretion rate of 10$^{-8}$ \ms~ yr$^{-1}$, a thermonuclear runaway would occur with a recurrence time of greater than 200 yr. At RT Cru's optical maximum, the accretion rate reaches 10$^{-8}$ \ms\, yr$^{-1}$, but this phase tends to only last for about 3-4 years. Around the time of the previous optical minimum, in 2007, the accretion rate was about 5.83$\times$10$^{-9}$ \ms\,yr$^{-1}$ (d/2.52 kpc)$^{2}$ \citep{2018A&A...616A..53L}, and during the optical minimum reported here, the accretion rate dropped to as low as 3.2$\times$10$^{-11}$ \ms\, yr$^{-1}$ (d/2.52 kpc)$^{2}$. Thus, the WD in RT~Cru accumulating material primarily during high states and taking more than a century to accumulate the mass needed to trigger a nova eruption is consistent with observations of high states only occurring every decade or so.

\begin{acknowledgements}

We acknowledge with thanks the variable star observations from the AAVSO International Database contributed by observers worldwide and used in this research. We thank Dr. Luis Mammana for his help during the HSH observing run at CASLEO. We acknowledge the use of public data from the \swift\ data archive as well as the frequent ToO observations performed upon our request. We thanks Dr. N. Schartel for the approval of our DDT request to observe RT~Cru with \xmm.

Based on observations obtained at the {\it Swope} and {\it Du Pont}, telescopes of the Las Campanas Observatory (through programs CN2012B-5, CN2013A-114, CN2019A-64). Based on observations obtained at the international Gemini Observatory (through the poor-weather program GS-2016B-Q-94), a program of NSF’s NOIRLab, which is managed by the Association of Universities for Research in Astronomy (AURA) under a cooperative agreement with the National Science Foundation. on behalf of the Gemini Observatory partnership: the National Science Foundation (United States), National Research Council (Canada), Agencia Nacional de Investigación y Desarrollo (Chile), Ministério de Ciencia, Tecnología e Innovación (Argentina), Ministério da Ci\^encia, Tecnologia, Inova\c{c}\~oes e Comunica\c{c}\~{o}es (Brazil), and Korea Astronomy and Space Science Institute (Republic of Korea). Based on data obtained at Complejo Astron\'omico El Leoncito (HSH-2021A1-01), operated under agreement between the Consejo Nacional de Investigaciones Cient\'ificas y T\'ecnicas de la Rep\'ublica Argentina and the National Universities of La Plata, C\'ordoba and San Juan. Based in part on data obtained with the Chiron Spectrograph, operated by the SMARTS partnership. 


AJP is a Consejo Nacional de Investigaciones Científicas y Técnicas fellow. GJML and NEN are members of the CIC-CONICET (Argentina) and acknowledge support from grant ANPCYT-PICT 0901/2017. NPMK acknowledges support from the UK Space Agency. JLS acknowledges support from SAO DD0-21118X and NASA award 80NSSC21K0715. R.A. acknowledges financial support from DIDULS PR\#1953853 by Universidad de La Serena. FMW acknowledges support from NSF grant AST-1614113. RLO acknowledges financial support from the Brazilian institution CNPq (PQ-312705/2020-4). YN acknowledges support by Bulgarian National Science Fund - project K$\Pi$-06-M58/1 and project K$\Pi$-06-H28/2.

\end{acknowledgements}

\bibliographystyle{aa}    
\bibliography{listaref_MASTER}

\begin{appendix}

\section{Optical spectroscopic and photometric series observing log}

\begin{table*}[ht!]
\caption{Log of optical SMARTS/Chiron spectra.}
\label{tab:chironlog}
\begin{tabular}{ccc}
\hline
Label & Date & Exposure time\\
 &  & [sec]\\
 \hline\hline
1 &	2019-01-26 &	2400 \\
2 &	2019-02-28 &	1800 \\
3 &	2019-03-06 &	1800 \\
4 &	2019-03-08 &	2400 \\
5 &	2019-03-17 &	600 \\
6 &	2019-03-19 &	600 \\
7 & 2019-03-21 &    928 \\
8 & 2019-05-24 &  600 \\
9 & 2019-07-26 & 900\\
10 & 2019-08-03 & 900\\
11 & 2019-08-01 & 1200\\
12 & 2019-12-01 & 1800\\
13 & 2020-02-05 & 1800\\
14 & 2020-03-06 & 1800\\
15 &	2020-12-13 &	1800 \\	
16 &	2021-01-17 &	1800 \\		
17 &	2021-02-17 &	1800 \\		
18 &	2021-03-15 &	1800 \\
19 &    2021-03-27 &    1800 \\
20 &	2021-04-11 &	1800 \\
21 &	2021-04-18 &	1800 \\
22 &	2021-05-01 &	1800 \\
23 &	2021-05-06 &	1800 \\
24 &	2021-05-15 &	1800 \\
25 &	2021-05-27 &	1200 \\
26 &	2021-06-17 &	1800 \\
27 &	2021-06-29 &	600 \\
28 &	2021-07-07 &	1800 \\
29 &	2021-07-22 &	600 \\
30 &	2021-08-28 &	600 \\
31 &    2022-04-15 &    600 \\

\hline
\end{tabular}
\end{table*}

\begin{table}
\begin{small}
\caption{Flickering measurements in $V$, $B$ and $U$ band. We show the date when the light curve was taken in MJD and calendar date; the amplitude calculated as {\it max-min} (see Section \ref{timing-aper}); the mean magnitude, band and the source of data. $TESS$ magnitudes were calibrated against contemporaneous $I$ AAVSO observations. The rows are sorted by filter. Data taken during the low accretion rate period are highlighted in bold-face. In comparison with other dates, the flickering amplitude during 2019 significantly decreased. 
}
\label{tab-flickering}
\begin{tabular}{cccccccc}
\hline\hline
MJD & Date & {\it max-min}& $\left\langle {\rm mag}\right\rangle $ & Band & Source & Obs. Lenght\tablefootmark{a} & t$_{exp}$\\
 &  & &  & &  & hrs & s\\
55691 & 2011-05-10 & 0.36 & 11.33 & V & AAVSO & 0.9 & 59 \\
55694 & 2011-05-13 & 0.66 & 11.30  & V & AAVSO & 2.0 & 60  \\
55952 & 2012-01-26 & 0.44 &  11.70 & V & AAVSO & 4.6 & 60 \\
56886 & 2014-08-17 & 0.65 &  11.97 & V & AAVSO & 3.5 & 60 \\
56893 & 2014-08-24 & 0.65 & 11.95 & V & AAVSO & 1.4 & 60  \\
57144 & 2015-05-02 & 0.20&  12.28 & V & AAVSO & 0.85 & 60  \\
57154 & 2015-05-12 & 0.16 & 12.26 & V & AAVSO & 0.9 & 60  \\
57158 & 2015-05-16 & 0.19 & 12.25 & V & AAVSO & 1.2  & 60  \\
57159 & 2015-05-17 & 0.22 & 12.19& V & AAVSO & 1.5 & 60  \\
57167 & 2015-05-25 & 0.19 & 12.25 & V & AAVSO & 1.0  & 60  \\
57170 & 2015-05-28 & 0.40 & 12.13 &  V & AAVSO  & 1.5 & 60 \\
57188 & 2015-06-15 & 0.21 & 12.12 &  V & AAVSO & 1.8  & 60  \\
57190 & 2015-06-17 & 0.19 &   12.10 &  V & AAVSO & 1.2 & 60  \\
57198 & 2015-06-25 &   0.24 & 12.26 &  V & AAVSO & 1.7  & 60  \\
57212 & 2015-07-09 & 0.24 & 12.18  &  V & AAVSO & 1.8  & 60  \\
57817 & 2017-03-05 & 0.28 & 12.75  &  V & AAVSO & 0.9  & 60 \\
57818 & 2017-03-06 & 0.42 & 12.72  &  V & AAVSO & 2.2  & 60 \\
58600 & 2019-04-27 & 0.002 & 8.933& TESS & TESS & 13 & 1800 \\
58615 & 2019-05-12 & 0.002 & 8.904 & TESS & TESS & 13 & 1800\\
59260	&	2021-02-15	&	0.09	&	12.643	&	V	&	HSH	& 2.4 & 60\\
59266	&	2021-02-21	&	0.18	&	12.643	&	V	&	HSH & 5.2 & 60\\
59307 & 2021-04-03 & 0.057 & 8.715 & TESS & TESS & 13 & 600\\
59311 &     2021-04-07 & 0.22 & 11.94 &   V & AAVSO & 1.3 & 60 \\
59312 &     2021-04-08 & 0.17 &  11.99 &   V & AAVSO & 3.9 & 60  \\
59317	&	2021-04-13	&	0.24	&	11.848	&	V	&	HSH	& 8.8 & 90\\
59318	&	2021-04-14	&	0.26	&	11.811	&	V	&	HSH	& 4.5 & 60\\
59320 & 2021-04-16 & 0.053 & 8.709 & TESS & TESS & 13 & 600\\
59333 & 2021-04-29 & 0.079 & 8.764 & TESS & TESS & 13 & 600\\
59347 & 2021-05-13 & 0.056 & 8.799& TESS & TESS & 13 & 600\\
59357	&	2021-05-21	&	0.20	&	11.802	&	V	&	HSH	& 3.8 & 90\\
59358	&	2021-05-22	&	0.16	&	11.720	&	V	&	HSH	& 2.6 & 60\\
59409	&	2021-07-12	&	0.21	&	11.522	&	V	&	HSH	& 1.47 & 45\\
59433	&	2021-08-06	&	0.14	&	11.570	&	V	&	HSH	& 2.42 & 60\\
59434	&	2021-08-07	&	0.19	&	11.610	&	V	&	HSH	& 2.32 & 40\\
59582	&	2022-01-03	&	0.23	&	11.304	&	V	&	HSH	& 2.6 & 30\\
59703	&	2022-05-03	&	0.16	&	11.374	&	V	&	HSH	& 2.18 & 60\\
\hline
57761	&	2017-01-08	&	0.48	&	16.319	&	U	&	Gemini	& 1.68 & 60\\
57829	&	2017-03-17	&	0.55	&	16.382	&	U	&	Gemini	& 1.2 & 60\\
57843	&	2017-03-31	&	0.62	&	17.106	&	U	&	Gemini	& 2.04 & 30\\
57858	&	2017-04-15	&	0.58	&	16.001	&	U	&	Gemini	& 3.11 & 60\\
58135	&	2018-01-17	&	0.32	&	17.699	&	U	&	Swope	& 4.54 & 180\\
{\bf 58523} 	&	{\bf 2019-02-09}	&	{\bf 0.07}	&	{\bf 15.893}	&	{\bf U}	&	{\bf DuPont}	& 4.03 & 240\\
{\bf 58694}	&	{\bf 2019-07-30}	&	{\bf 0.08}	&	{\bf 17.763}	&	{\bf U}	&	{\bf Swope} & 1.7 & 300	\\
\hline
{\bf 58520}	&	{\bf 2019-02-06}	&	{\bf 0.02}	&	{\bf 12.710}	&{\bf B}	&{\bf DuPont}	& 2.4 & 60\\
{\bf 58522}	&	{\bf 2019-02-08}	&	{\bf 0.02}	&	{\bf 13.388}	&{\bf B}	&{\bf DuPont}	& 3.96 & 60\\
{\bf 58544}	&	{\bf 2019-03-02}	&	{\bf 0.02}	&	{\bf 14.865}	&{\bf B}	&{\bf Swope}	& 3.52 & 90\\
{\bf 58545}	&	{\bf 2019-03-03}	&	{\bf 0.02}	&	{\bf 14.408}	&{\bf B}	&{\bf Swope}	& 4.08 & 90 \\
59260	&	2021-02-15	&	0.25	&	14.026	&	B	&	HSH	& 2.4 & 120\\
59260	&	2021-02-16	&	0.23	&	14.026	&	B	&	HSH	& 1.6 & 120\\
59266	&	2021-02-21	&	0.37	&	14.026	&	B	&	HSH	& 5.85 & 120\\
59317	&	2021-04-13	&	0.35	&	12.817	&	B	&	HSH	& 9.71 & 120\\
59318	&	2021-04-14	&	0.36	&	12.771	&	B	&	HSH	& 4.53 & 120\\
59357	&	2021-05-21	&	0.34	&	12.631	&	B	&	HSH	& 3.82 & 120\\
59358	&	2021-05-22	&	0.21	&	12.482	&	B	&	HSH	& 2.58 & 120\\
59409	&	2021-07-12	&	0.33	&	12.298	&	B	&	HSH	& 1.53 & 120\\
59433	&	2021-08-06	&	0.23	&	12.434	&	B	&	HSH	& 2.48 & 150\\
59434	&	2021-08-07	&	0.27	&	12.474	&	B	&	HSH	& 2.31 & 40\\
59582	&	2022-01-03	&	0.31	&	12.278	&	B	&	HSH	& 2.65 & 60\\
59703	&	2022-05-03	&	0.22	&	12.161	&	B	&	HSH	& 2.18 & 90\\

\hline
\end{tabular}
\end{small}
\tablefoottext{a}{Listed in days for $TESS$ observations}
\end{table}  

\begin{figure*}
\begin{center}
\includegraphics[scale=1]{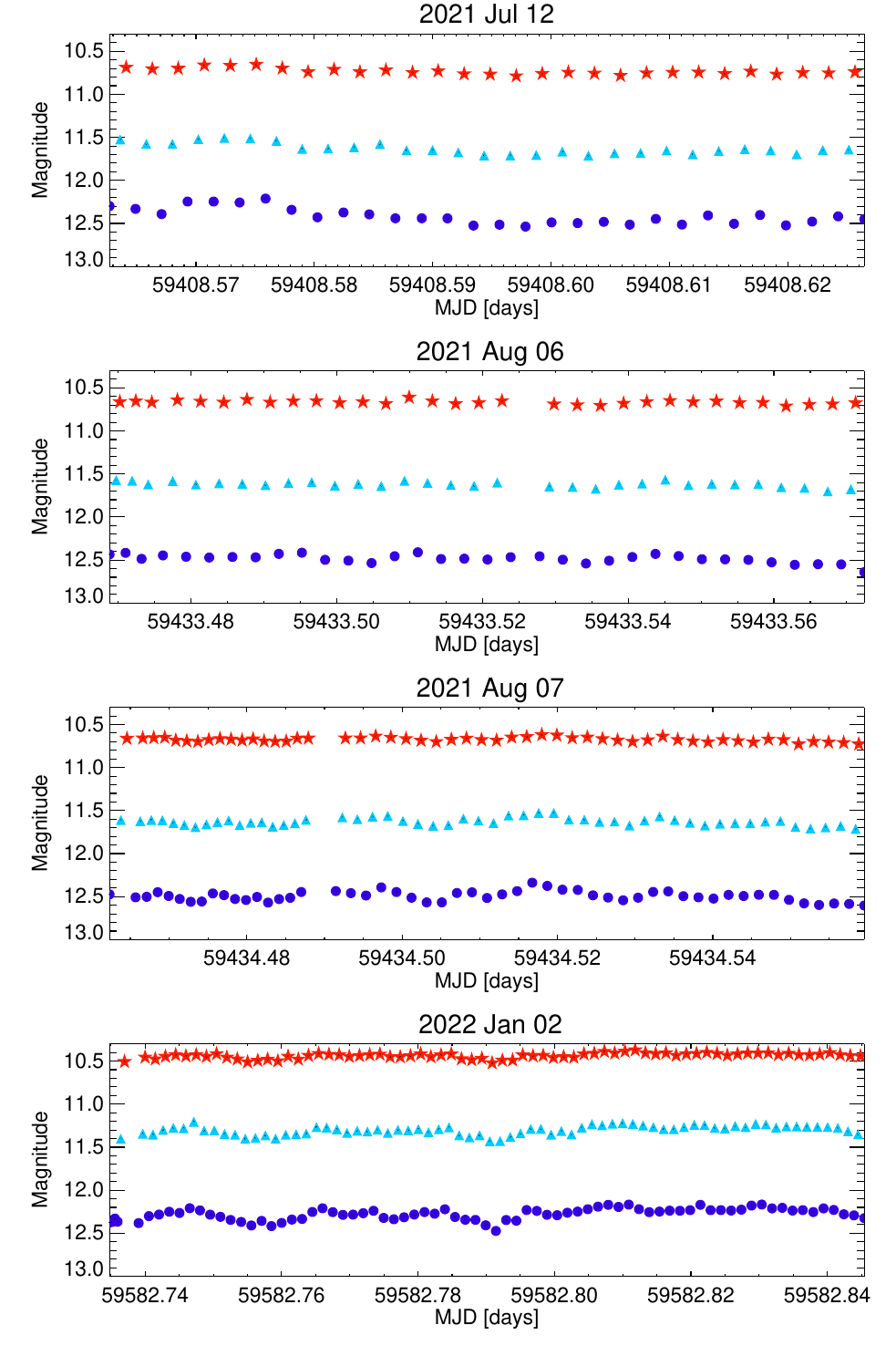}
\caption{Light curves of our short-term optical monitoring of RT~Cru, obtained with the $HSH$ telescope. R (red stars), V (light blue triangles) and B (blue circles) magnitudes are displayed, while error bars are smaller than the symbols. Cadence and observing length are listed in Table ~\ref{tab-flickering}}.
\label{fig.rtcru.hsh}
\end{center}
\end{figure*}

\begin{figure*}
\begin{center}
\includegraphics[scale=1]{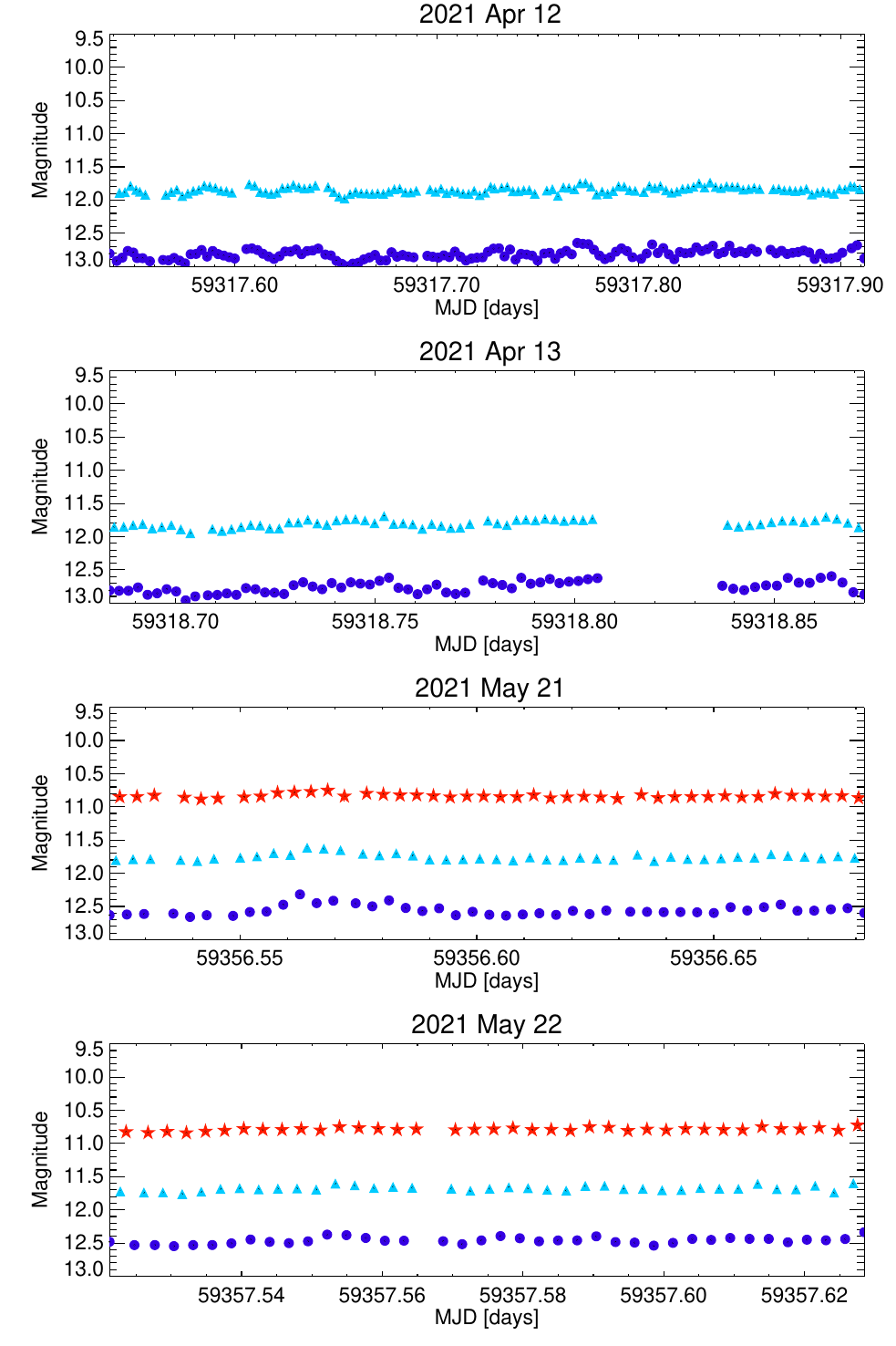}
\caption{{\it Same as Fig. \ref{fig.rtcru.hsh}}.
}
\label{fig.rtcru.hsh2}
\end{center}
\end{figure*}

\begin{figure*}
\begin{center}

\includegraphics[scale=1]{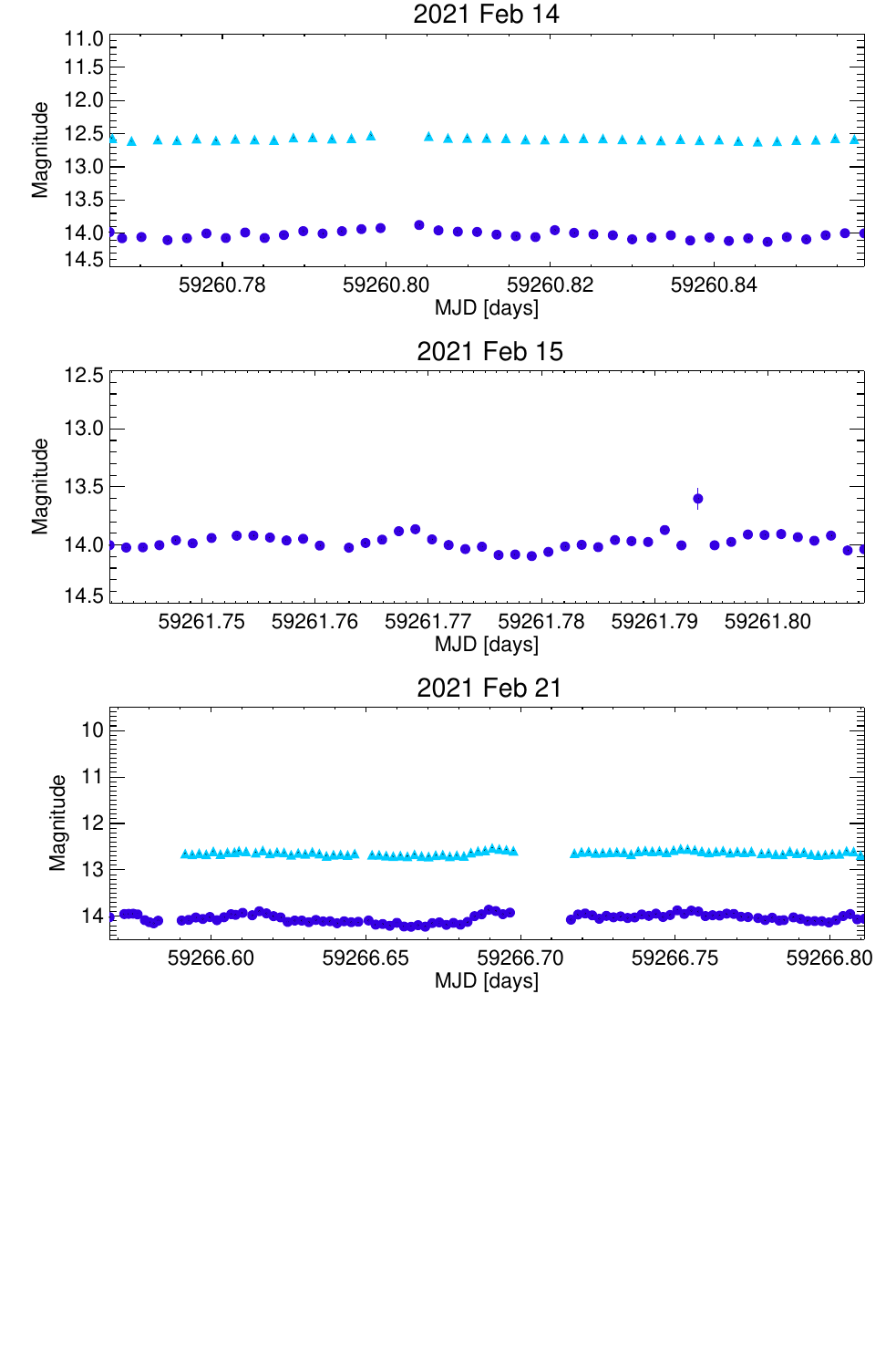}
\caption{{\it Same as Fig. \ref{fig.rtcru.hsh}}..
}
\label{fig.rtcru.hsh3}
\end{center}
\end{figure*}

\begin{figure*}
\begin{center}
\includegraphics[scale=1]{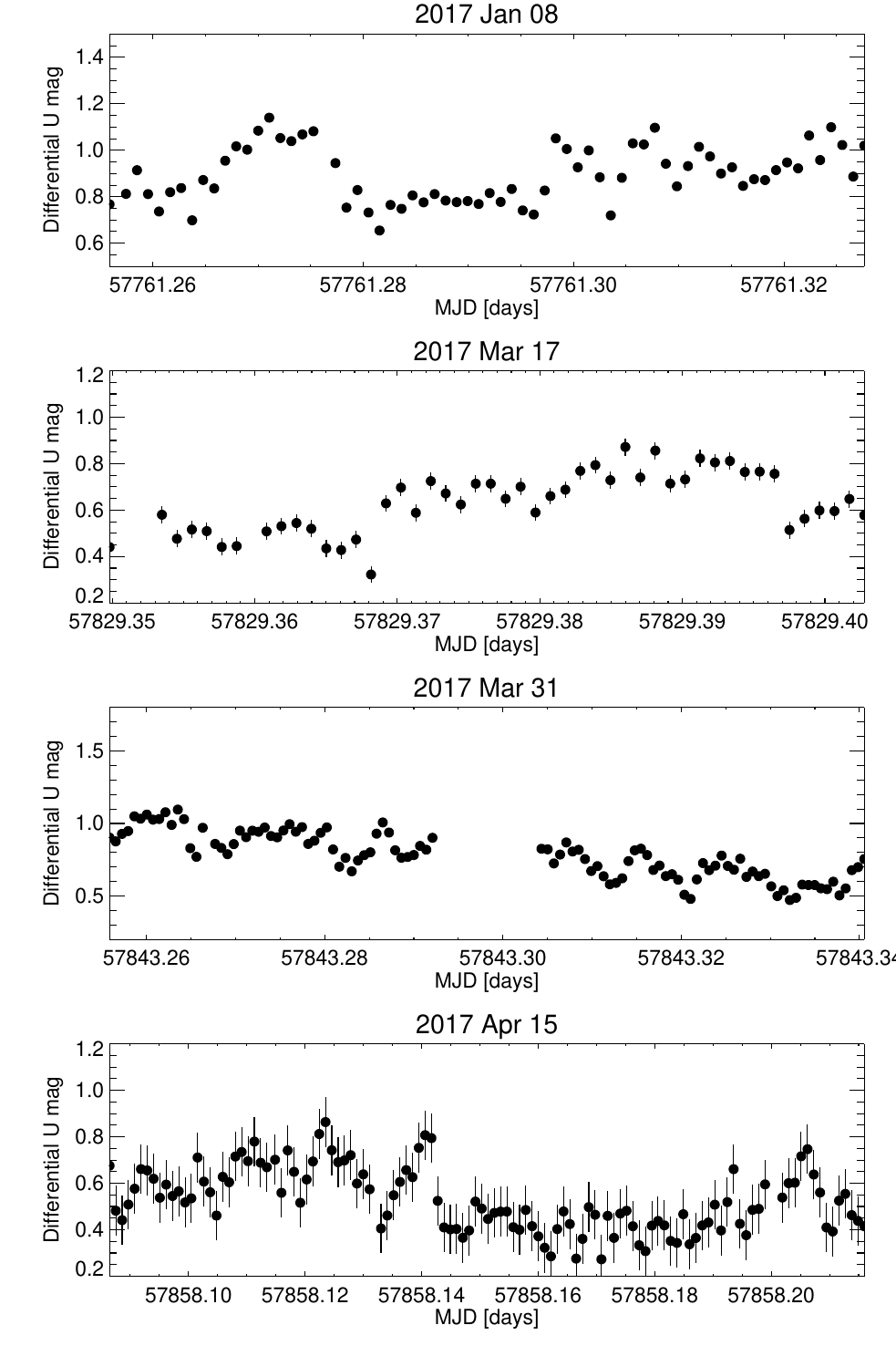}
\caption{Light curves of our short-term optical monitoring of RT~Cru, obtained with the $Gemini$-South telescope. (see Table ~\ref{tab-flickering}).
}
\label{fig.rtcru.gemini}
\end{center}
\end{figure*}

\begin{figure*}
\begin{center}
\includegraphics[scale=1]{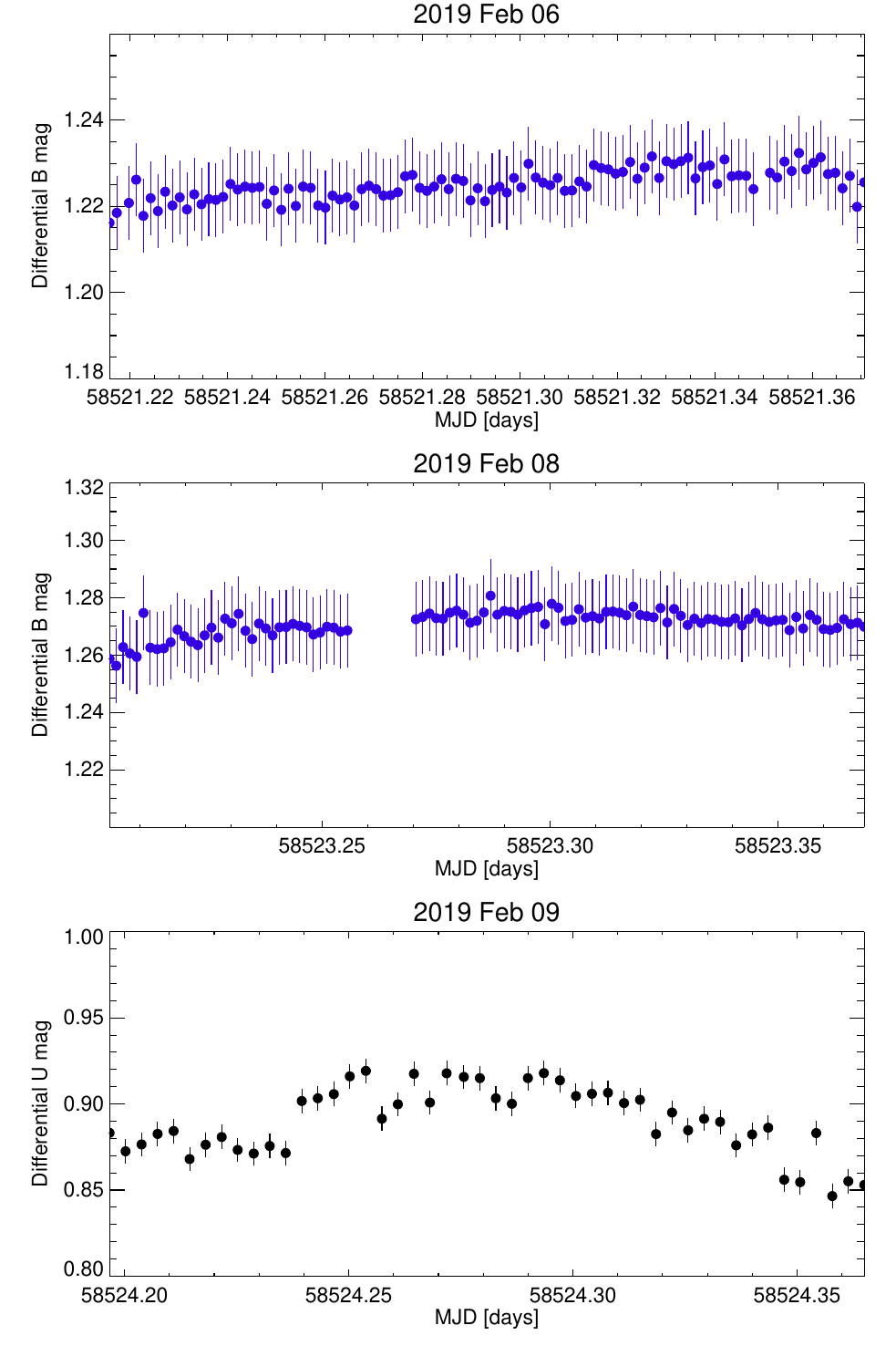}
\caption{Light curves of our short-term optical monitoring of RT~Cru, obtained with the $DuPont$ telescope. Blue circles show B-band differential magnitudes while black circles show U-band magnitudes (see Table ~\ref{tab-flickering}).
}
\label{fig.rtcru.dupont}
\end{center}
\end{figure*}

\begin{figure*}
\begin{center}
\includegraphics[scale=1]{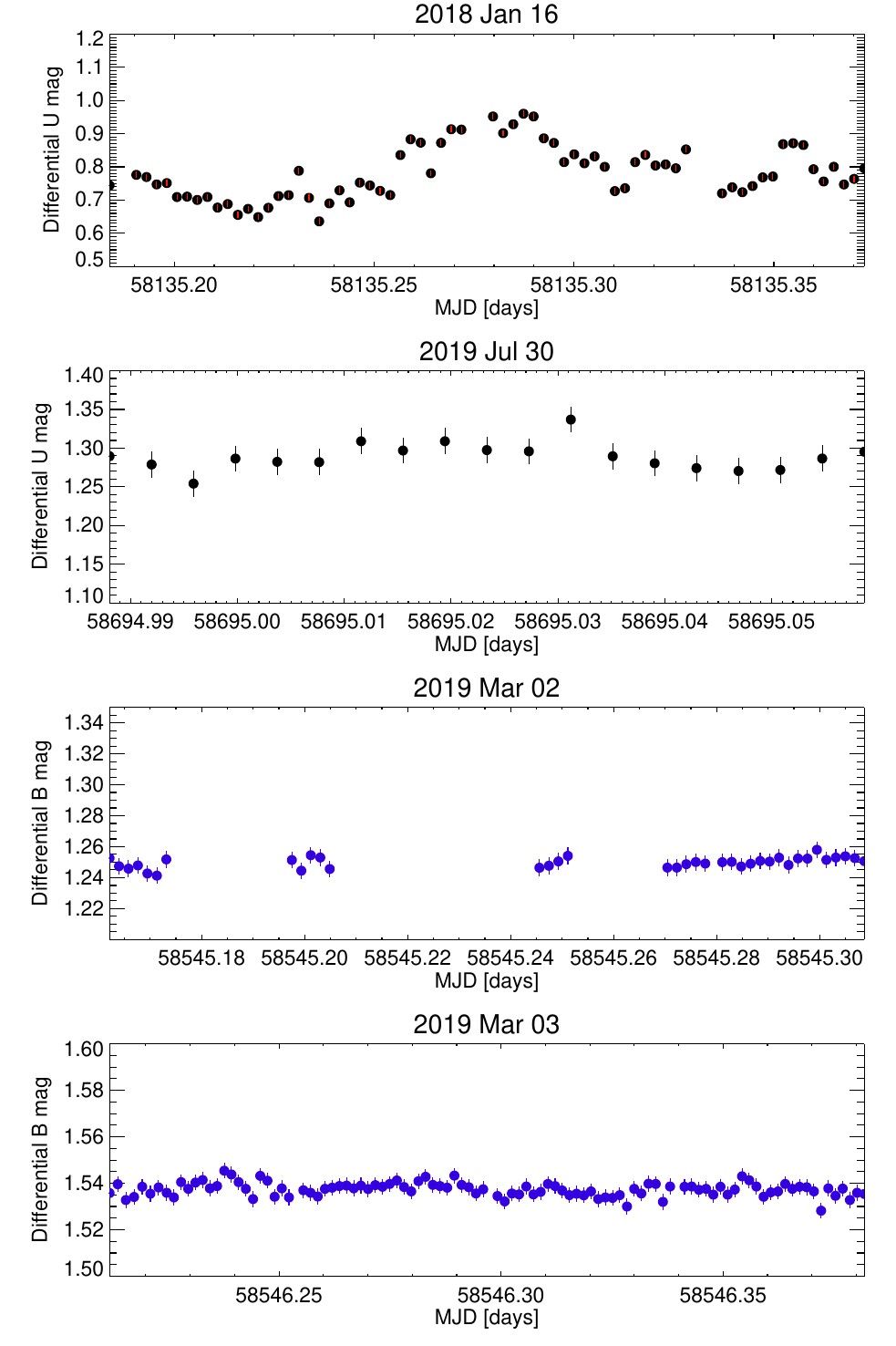}
\caption{Light curves of our short-term optical monitoring of RT~Cru, obtained with the $Swope$ telescope. Blue circles show B-band differential magnitudes while black circles show U-band magnitudes (see Table ~\ref{tab-flickering}).
}
\label{fig.rtcru.swope}
\end{center}
\end{figure*}

\begin{figure*}
\begin{center}
\includegraphics[scale=0.9]{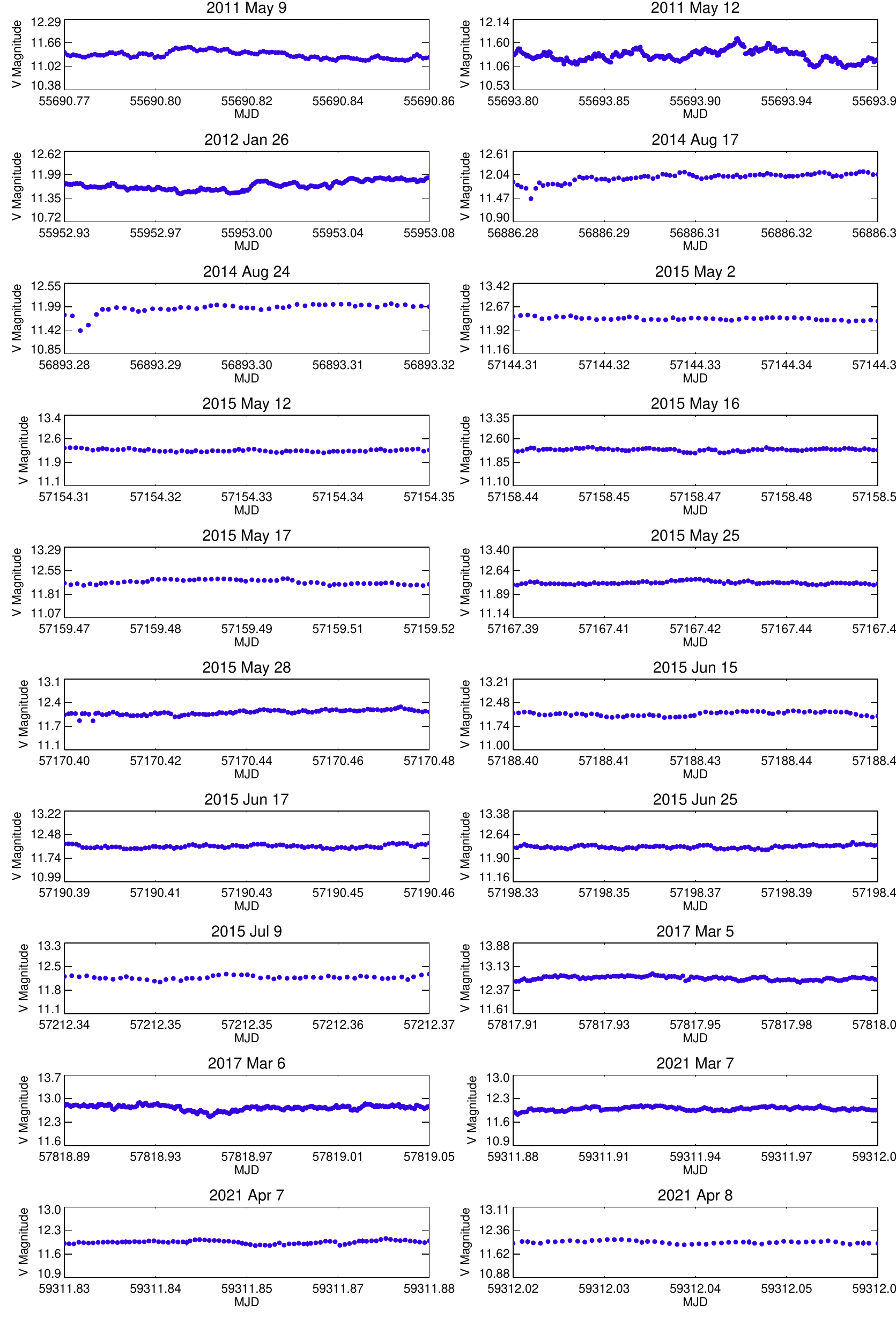}
\caption{$V$-band light curves of RT~Cru, obtained from the AAVSO database (see Table ~\ref{tab-flickering}).
}
\label{fig.rtcru.aavso}
\end{center}
\end{figure*}
\end{appendix}
\end{document}